\documentclass[prfluids,showpacs,notitlepage,nofootinbib,longbibliography,aps]{revtex4-2}

\usepackage[utf8]{inputenc}
\usepackage{epsf,graphicx}
\usepackage{natbib}
\usepackage{multirow}
\usepackage{amsmath,gensymb,amssymb}
\usepackage{siunitx}
\usepackage[bottom]{footmisc}
\usepackage{lipsum}
\usepackage{dcolumn}
\usepackage{xcolor}
\usepackage{upgreek}
\usepackage{float,ulem}

\usepackage{bm}

\usepackage{hyperref}
\hypersetup{
colorlinks=true, 
citecolor=magenta,
breaklinks=true, 
urlcolor= blue, 
linkcolor= blue, 
bookmarksopen=true, 
}
\usepackage{natbib}
\usepackage{color}

\begin{document}

\title{Transition from wave turbulence to acousticlike shock-wave regime}	

\author{Guillaume Ricard}
\email{guillaume.ricard@u-paris.fr}
\affiliation{Universit\'e Paris Cité, CNRS, MSC, UMR 7057, F-75013 Paris, France}

\author{Eric Falcon}
\email{eric.falcon@u-paris.fr}
\affiliation{Universit\'e Paris Cité, CNRS, MSC, UMR 7057, F-75013 Paris, France}

\begin{abstract}
We report on the experimental observation of a transition from a dispersive wave turbulence regime to a nondispersive regime involving shock waves on the surface of a fluid. We use a magnetic fluid in a canal subjected to an external horizontal magnetic field to tune the dispersivity of the system. For a low magnetic field, gravity-capillary wave turbulence is observed, whereas for a high enough field, random steep coherent structures arise which are found to be shock waves. These shock waves create singularities in the second-order difference of the surface elevation, leading to an $\omega^{-4}$ frequency power spectrum. This spectrum is also found to be controlled by the number and amplitude of the shocks and is well captured by a model based on a random Dirac-$\delta$ distribution (Kuznetsov-like spectrum). Finally, the shock-amplitude statistics exhibits a power-law distribution with an exponent close to the predictions of the one-dimensional random-forced Burgers equation. This shock-wave regime, discovered here for surface waves, thus paves the way to better explore their properties.
\end{abstract}

\maketitle


\section{Introduction}
Wave turbulence is a statistical state in which numerous random weakly nonlinear waves interact with each other. This phenomenon is described by the weak-wave turbulence theory (WTT) which predicts a power-law cascade of the wave energy spectrum from large to small scales~\cite{ZakharovBook,NazarenkoBook,Newell2011}. This out-of-equilibrium stationary state occurs in various domains with different scales such as ocean surface waves, plasma waves, hydroelastic waves, elastic waves on a plate, internal or inertial waves on rotating stratified fluids, and optical waves~\cite{NazarenkoBook}. Despite its success in predicting analytically the wave spectrum, WTT requires many assumptions (e.g., infinite system, weak nonlinearity, constant energy flux, timescale separation, and dispersive waves), which can be difficult to satisfy experimentally. Although wave turbulence has been assessed in different experimental systems~\cite{FalconPRL07,laurie2012,Deike2013,HassainiPRE2019,monsalve2020,ARFM2022}, it is of paramount interest to know the validity domain of the theory in experiments regarding its assumptions. For example, finite-size effects are beginning to be considered theoretically~\cite{lvov2006,BanksPRL2022} and experimentally~\cite{Issenmann2013,deike2015,hassaini2018,CazaubielPRL2019} for hydrodynamics surface waves. Finite-amplitude effects have also been tackled to address the existence of a transition from weak to strong wave turbulence~\cite{NazarenkoBook}.

In comparison, few studies have investigated whether or not wave turbulence exists in a nondispersive wave system. In this case, waves of different frequencies travel with the same phase velocity and thus cannot transfer energy between each other by resonant interactions~\cite{NazarenkoBook}. This leads to the breaking of a main assumption of WTT, and coherent structures such as solitons or shocks are thus expected due to cumulative effects of the nonlinearity~\cite{NewellJFM1971,LvovPRE1997}. This has been the source of a long-standing debate about whether acoustics waves should be considered as a random set of shocks (leading to the Kadomtsev-Petviashvili spectrum)~\cite{kadomtsev1973} or if WTT is applicable for their description~\cite{Zakharov1970}. Indeed, three-dimensional acoustic WTT could be theoretically possible because the large range of possible wave directions in three dimensions acts as an effective dispersion~\cite{NazarenkoBook,NewellJFM1971,LvovPRE1997,Zakharov1970}, although yet unsupported by a rigorous proof~\cite{GriffinPRL2022}. Conversely, WTT is not applicable for two-dimensional (2D) nondispersive acoustic waves, but can be regularized by weakly dispersive effects leading to predictions for 2D weakly dispersive acoustic wave turbulence~\cite{GriffinPRL2022}. Weakly dispersive wave turbulence also occurs theoretically or numerically for Alfv\'en waves in plasma~\cite{galtier2000}, gravitational waves in the early universe~\cite{GaltierPRL2017} and elastic waves on a stretched membrane~\cite{HassainiPRE2019}. Experimentally, a weakly dispersive wave regime can be obtained on the surface of a magnetic fluid subjected to an external horizontal magnetic field. The latter modifies the dispersion relationship of surface waves adding a nondispersive term that is tunable experimentally~\cite{RosensweigBook}. In this case, dispersive wave turbulence is evidenced experimentally in two dimensions because of the anisotropic dispersion relation, nondispersivity occurring only in the magnetic field direction~\cite{DorboloPRE2011,Kochurin2022}. Another method to experimentally control the wave dispersion is to decrease the fluid depth of gravity-capillary wave turbulence from a deep regime to a shallow one~\cite{FalconEPL2011,HassainiPRF2017}. This deep-to-shallow transition leads to a less steep gravity wave spectrum, the formation of a depth-dependent hump in the capillary spectrum (as an analog of a bottleneck effect) for a weak forcing~\cite{FalconEPL2011}, and the formation of coherent structures as solitons when the forcing is strong enough~\cite{HassainiPRF2017,ZakharovPR2004}. 

Here, we use a one-dimensional (1D) canal filled with a magnetic fluid subjected to an external horizontal magnetic field to tune the dispersivity of the wave system within a deep-water regime. At a low magnetic field, the classical quasi-1D dispersive gravity-capillary wave turbulence is observed~\cite{Ricard2021}, whereas at a high enough field a nondispersive regime is reached. In the latter, we observe the emergence of random shock waves, keeping their shape over time, with a very steep profile close to the one derived from the 1D Burgers equation~\cite{burgers1948}, although not reaching a fully vertical front. They are characterized by a discontinuity that leads to a Dirac-$\delta$ singularity in the second-order difference of their amplitude. We show that these shock waves are coherent structures rich in the frequency domain, which carry energy over the canal. They thus become the main mechanism building the wave energy spectrum. Indeed, we found that the energy spectrum of these shocks agrees with a model of a Kuznetsov-like spectrum of second-order singularities~\cite{KuznetsovJETP2004}. The shock-wave statistics are also reported and show that their probability distribution is close to the one of a diluted gas of shocks driven by the 1D random-forced Burgers equation~\cite{Chekhlov1995,weinan1997,E1999,Bec2007,Frisch2001}. A phase diagram of the wave turbulence and shock-wave regimes is also reported as a function of the control parameters. The energy transfer driven by the shock waves is thus fundamentally different from the local one occurring in wave turbulence by nonlinear wave resonant interactions.

The article is organized as follows. We first present in Sec.~\ref{DispRel} some theoretical background (dispersion relationship, magnetic steepening, and energy spectrum predictions). Section~\ref{Expsetup} presents the experimental setup. Section~\ref{ExpResults} shows the experimental results on the wave energy spectrum (using spatiotemporal, time-frequency, and frequency analyses), the energy flux, and timescales. Section~\ref{ShockStats} focuses on the nondispersive regime emphasizing the presence of dissipative coherent structures as shock waves, and their statistics. Section~\ref{DisccusSpectra} presents the model used to predict the shock wave spectrum and the conditions for an experimental agreement. We summarize in Sect.~\ref{Conc}.

\section{Theoretical background}\label{DispRel}
\subsection{Dispersion relation}
\begin{figure}[h!]
    \centering
    \includegraphics[width=0.5\linewidth]{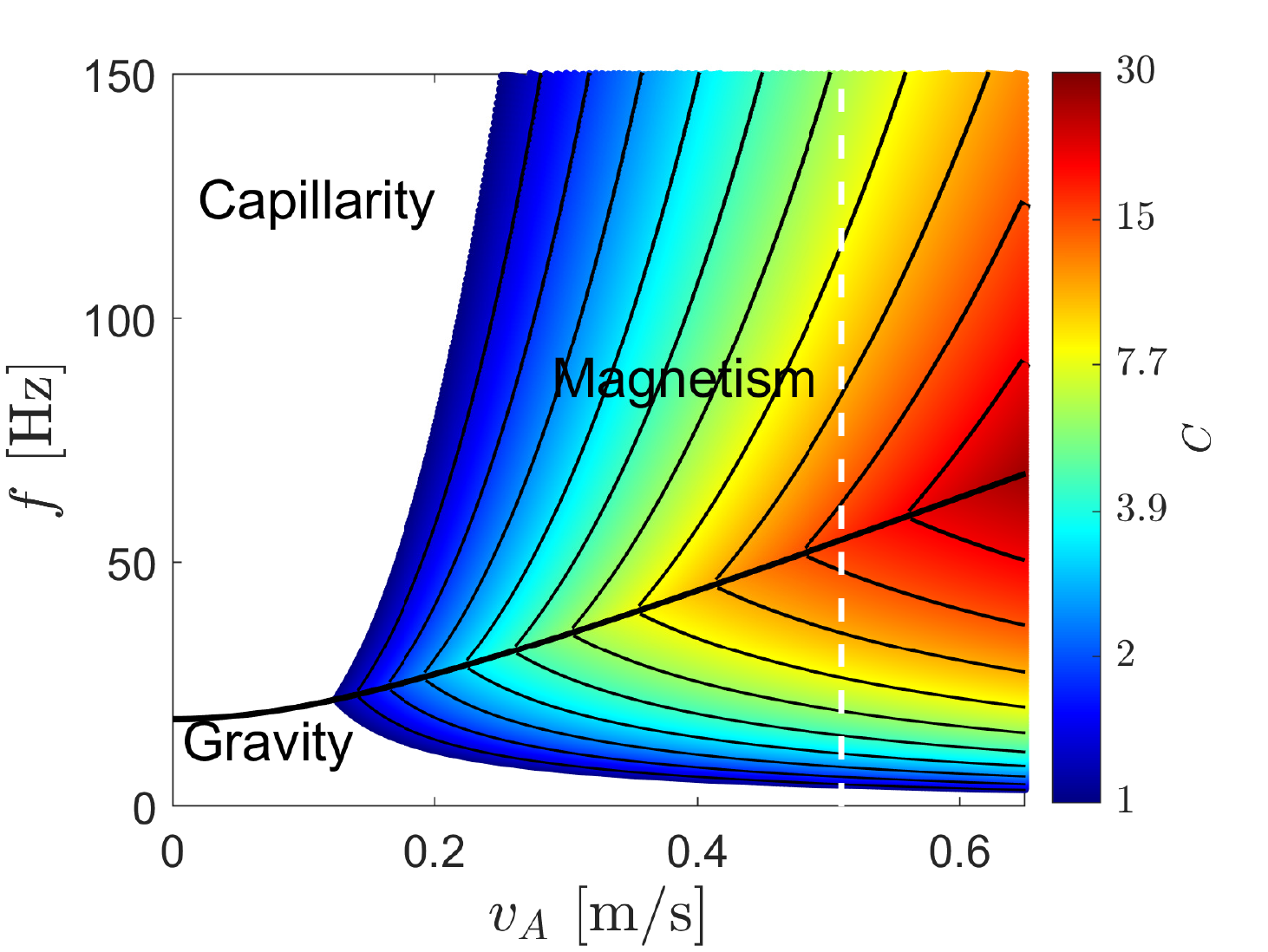} 
    \caption{Theoretical diagram of the predominance of the gravity, capillary, and magnetic regimes. Here, $C$ is defined as how much the magnetic term is bigger than the two others. The experimental ranges are $f<100$~Hz and $v_A< 0.55$ m/s (i.e., $B<760$~G). The white vertical dashed line corresponds to the run at $v_A=0.51$ m/s.}
    \label{fc_vA}
\end{figure}

The dispersion relation of one-dimensional linear deep-water inviscid gravity-capillary waves reads $\omega^2=gk+(\gamma/\rho)k^3$, with $\omega=2\pi f$ the angular frequency, $k$ the wave number, $g$ the acceleration of gravity, $\gamma$ the surface tension, and $\rho$ the density of the liquid~\cite{Lamb1932}. For a magnetic liquid subjected to a horizontal magnetic induction $B$ (collinear to the wave propagation), an additional nondispersive term, i.e., acousticlike term in $\omega \sim k$, has to be taken into account for which its strength is controlled by $B$. The corresponding dispersion relation then reads \cite{RosensweigBook,Zelaco1969}
\begin{equation}
    \omega^2=gk+\frac{\gamma}{\rho}k^3+v_A^2(B)k^2,
    \label{relat_disp}
\end{equation}
where $v_A^2=\frac{\mu_0M^2}{1+\mu/\mu_0}$ is the characteristic nondispersive velocity analogous of the Alfv\'en wave velocity in plasma~\cite{Alfven1942}, $M(B)$ is the magnetization within the liquid depending on the applied magnetic field induction $B$, $\mu_0=4\pi\times10^{-7}$~Tm/A is the magnetic permeability of a vacuum, and $\mu=\mu_0(1+\frac{\partial M}{\partial B})$ is the liquid permeability~\cite{RosensweigBook}. Note that $B$ should not be confused with the external magnetic field, $H=B/\mu-M$, even if $B$ will be hereafter referred to as the magnetic field. The dispersion relation can be rewritten as
\begin{equation}
    \omega=v_A(B)k\sqrt{1+\alpha k^{-1}+\beta k},
    \label{relat_disp2}
\end{equation}
with $\alpha=g/v_A^2$ and $\beta=\gamma/(\rho v_A^2)$. A nondispersive regime $\omega \sim k$ could be obtained if the gravity and the capillary terms are much smaller than the magnetic one, i.e., if
\begin{equation}
         \alpha k^{-1}<C\ \ \ {\rm and}\ \ \ \beta k<C\ ,
    \label{dominence}
\end{equation}
where $C$ is a chosen constant quantifying the ratio between the magnetic term and the gravity or capillary one. Using the dispersion law of Eq.~\eqref{relat_disp}, we plot in Fig.~\ref{fc_vA} the theoretical diagram of the predominance of the gravity, capillary, and magnetic regimes~\cite{DorboloPRE2011} as a function of the parameter $C$. From our ranges of experimental parameters used afterward, we can reach $C\sim20$, i.e., a magnetic term larger than 20 times each of the other two. This will be possible because of the use of a ferrofluid with a high magnetic susceptibility and a relatively low viscosity (see below).

\subsection{Magnetic wave steepening}\label{MagneticEffect}
It is worth noting that in the dispersion law of Eq.~\eqref{relat_disp} the magnetic term comes from the spatiotemporal fluctuations of the magnetic field generated at the liquid-gas wavy interface to satisfy the magnetic boundary conditions at the interface~\cite{RosensweigBook}.
The magnetic fluctuations $h$ at the interface in the direction $Ox$ of the constant horizontal field $H$ are obtained due to a calculation similar to the one performed in~\cite{RosensweigBook} for a vertical magnetic field and read $h_{1}=h_2=M\eta k(1+\mu/\mu_0)$,
where indices 1 and 2 refer to the magnetic liquid and the gas, respectively. The more important the surface perturbation is, the more the fluctuation in the magnetic field appears. With typical values used here ($B\approx760$~G, $\mu_0 M\approx340$~G, $\mu/\mu_0\sim1.05$, $k\approx500$~m$^{-1}$, and $\eta\approx\pm1$~mm), the magnetic induction fluctuations $b_1=\mu h_1$ and $b_2=\mu_0h_2$ are about $\pm80$~G, that is to say, about $\pm10\%$ of the applied value. We can thus infer the magnetic force $F_m$ acting on the fluid in the $x$ direction as $F_m=\frac{\mu_0 M}{\rho}\frac{\partial h}{\partial x}=\rho v_A^2k^2\eta$. $F_m$ acts more at the extrema of a wave than at its base ($\eta=0$) and thus leads to a steepening of the wave and a difference of the fluid velocity along the wave height. This mechanism is the source of the appearance of shock waves as it is for the Burgers shock waves~\cite{burgers1948} (see Sec.~\ref{signal}). Note that no experimental comparison is performed here to check the above theoretical predictions on the field fluctuations $h$, but such a comparison is done to explain qualitatively the physical process of the shock-wave formation observed below. Note also that magnetic stress, called Maxwell stress, occurs at the interface of a magnetic fluid~\cite{RosensweigBook}. For a horizontal magnetic field, this stress $s_n=-\frac{1}{2}\mu_0H^2$, normal to the surface, tends to flatten the surface wave acting as a stabilizer. This higher-order effect will be not visible here but might appear at higher $v_A$, although not achievable experimentally.

\subsection{Energy spectra}
Wave turbulence arises from the interaction of weakly nonlinear waves and is described by the weak turbulence theory~\cite{ZakharovBook,NazarenkoBook}. The latter predicts that the wave energy spectrum follows a power-law cascade of the scale (frequency or wavenumber) only for a system involving a single term in its dispersion relation $\omega(k)$. For example, in one dimension, pure gravity waves dominated by five-wave resonant interactions are predicted to have a power spectrum of the surface elevation $\eta$ as $S_\eta \sim \omega^{-17/4}$~\cite{Dyachenko1995five}. It has been also observed experimentally that 1D capillary waves are dominated by five-wave resonant interactions and follow a power spectrum in $S_\eta\sim\omega^{-31/12}$~\cite{Ricard2021}. Thus, for a 1D gravity-capillary system (with no magnetic field), these two asymptotic spectra are thus expected, the pure gravity spectrum for large enough scales ($f\lesssim 5$~Hz) and the pure capillarity one for small enough scales ($f\gtrsim 50$~Hz)~\cite{ARFM2022}. However, the finite size of our experimental system and the nonvanishing viscosity of the fluid used here will lead to work in the intermediate-frequency scales and thus to an entanglement of the gravity and capillary effects~\cite{ARFM2022}. Indeed, for a 1D gravity-capillary system, we previously reported experimentally a power-law spectrum in $S_\eta\sim\omega^{-3.3\pm0.2}$ in the intermediate-scale range as a result of the occurrence of three-wave interactions ~\cite{Ricard2021} [see also the purple curve in Fig.~\ref{Spectromeg}(a)].

Coherent structures are more likely to appear in one dimension than in higher dimensions~\cite{ZakharovPR2004}. For instance, transitions from wave turbulence to solitonic regimes have been predicted theoretically~\cite{ZakharovPR2004} and observed experimentally~\cite{HassainiPRF2017} for 1D gravity waves in shallow water, coherent structures such as Korteweg--de Vries solitons occurring as a result of the weak dispersion. For 1D deep-water gravity waves, other types of solitons, e.g., Peregrine solitons or envelope solitons, were observed experimentally~\cite{CazaubielPRF2018,MichelPRF2020}, but are not expected in our study. Nevertheless, since our system is nondispersive at high $v_A$, other coherent structures could arise such as singularities~\cite{ZakharovPR2004}. Singularities can be defined by local discontinuities of order $n$ in the wave field, i.e., leading to a Dirac-$\delta$ distribution on the $n$th-order derivative of the wave field $\partial^n\eta$. As discontinuities contain energy at all frequency scales~\cite{KuznetsovJETP2004,NazarenkoJFM2010}, these coherent structures would lead to a spectrum only driven by their geometry, i.e., the order of the discontinuity. Since the power spectrum of a Dirac-$\delta$ distribution occurring on $\partial^n\eta$ is a white noise, i.e., $S_{\partial^n{\eta}}\sim$ const, one has thus, by integration, the spectrum of $\eta$  in $S_\eta\sim\omega^{-2n}$. For discontinuities of the first order $n=1$ (e.g., shock waves in the Burgers' equation) an acoustic spectrum in $S_\eta\sim\omega^{-2}$ is thus excepted, i.e., the Kadomtsev-Petviashvilli spectrum~\cite{Saffman71,kadomtsev1973,KuznetsovJPP08}. If discontinuities are of second order $n=2$, e.g., sharp-crested waves, or shock waves not reaching a fully vertical front, one thus expects to obtain a spectrum in $S_\eta\sim\omega^{-4}$ (or Kuznetsov-like spectrum)~\cite{KuznetsovJETP2004}. 

\begin{figure}[h!]
    \centering
    \includegraphics[width=0.5\linewidth]{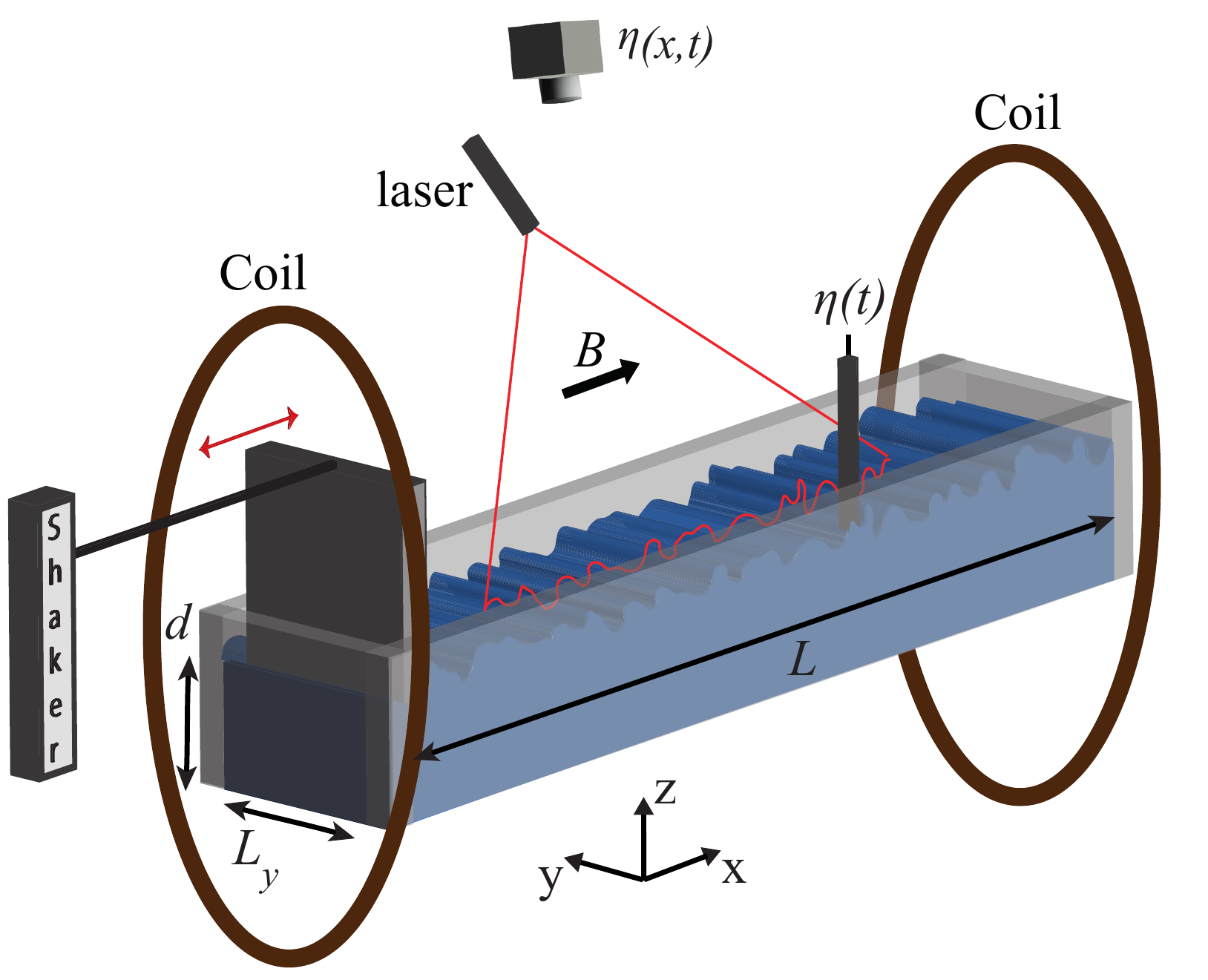} 
    \caption{Experimental setup. A pair of Helmholtz coils generates a horizontal homogeneous magnetic field $B$ on the ferrofluid surface. Random waves are driven by a wave maker linked to a shaker at one end of the canal. The wave elevation $\eta(t)$ is measured at a single point using a capacitive wire gauge, and resolved in space and time $\eta(x,t)$ with a laser sheet profilometry using a camera and a laser sheet illuminating a horizontal line of the free surface.}
    \label{setup}
\end{figure}

\section{Experimental setup}\label{Expsetup}
Experiments were performed in a canal made of polytetrafluoroethylene, i.e., Teflon, to decrease the wetting, with a length $L=15$~cm and a width $L_y=2$~cm (see Fig.~\ref{setup}). This hydrophobic canal is filled up to a depth $d=2$~cm with a ferrofluid (see below). A shaker linked to a wave maker is located at one end to inject energy in a narrow random frequency bandwidth $f_0\pm\Delta F$, with $f_0=8.5$~Hz and $\Delta F=2.5$ Hz. Since $L\gg L_y$, waves propagate only in the longitudinal $(Ox)$ direction and are thus considered to be quasi-1D~\cite{Ricard2021}. The whole setup is located between two vertical coils in Helmholtz configuration, 25 cm in internal diameter, generating a horizontal magnetic field ($B\in[0,800]$~G) homogeneous on the liquid surface. Two measurement methods of surface elevation are used: a single point measurement and a laser sheet profilometry (LSP). The temporal variations of the surface elevation $\eta(t)$ are measured at a single point using a homemade capacitive wire gauge (0.22~mm in diameter and 10~$\upmu$m vertical resolution)~\cite{FalconPRL07} with a 2 kHz sampling frequency leading thus to a resolved frequency up to 1~kHz and thus to a discretization time $dt=0.5$~ms. A space- and time-resolved wave-field measurement $\eta(x,t)$ is reached by the LSP method. A camera (Basler, 200 frames/s) is located above the canal and the wave field is illuminated over 8 cm with a laser sheet at an angle of $\alpha=45^\circ$ with respect to the horizontal (see Fig.~\ref{setup}). The horizontal shift $\Delta y(x,t)$ of the laser sheet along $Oy$ detected by the camera is hence directly linked to the surface elevation by $\eta(x,t)=\Delta y(x,t)/\tan{(\alpha)}=\Delta y(x,t)$~\cite{Aulnette2019}. The horizontal and vertical resolutions of LSP are 43~$\upmu$m. The wave elevation is monitored for both measurements for $\mathcal{T}=15$~min.

We use a Ferrotec PBG400 ferrofluid. This black-brown opaque ferrofluid offers high magnetization, high colloidal stability, and superparamagnetic properties. It is a water-based (with polyethylene glycol) suspension synthesized with 7.9\% by volume of ferromagnetic particles (Fe$_3$O$_4$ iron oxide, 10~nm in diameter). The properties of the liquid are density $\rho=1400$~kg/m$^3$, surface tension $\gamma=34$~mN/m, kinematic viscosity $\nu = 2.86\times 10^{-6}$~m$^2$/s, magnetic saturation $M_{sat}=440$~G, and initial susceptibility $\chi_i=3.28$. Note that $M_{sat}=\lim\limits_{B \to \infty}M$ and $\chi_i=\frac{\partial M}{\partial B}|_{B=0}$ are obtained due to the magnetization curve $M(B)$ provided by Ferrotec. Here $M(B)$ is also used to compute the characteristic velocity $v_A(B)$ used in Eq.~\eqref{relat_disp} (see Appendix~\ref{Ferro}). The ferrofluid high sensibility to magnetic effects with a relatively low viscosity is crucial to reach experimentally a significant inertial range (see Fig.~\ref{fc_vA}). To quantify nonlinearities, we measure the wave steepness as $\epsilon\equiv\sigma k_{m}$, where $\sigma$ is the standard deviation of the surface elevation signal, computed as $\sqrt{\overline{\eta(t)^2}}$ or $\overline{\sqrt{ \int_L\eta(x,t)^2dx/L}}$ (the overline is time average), and $k_{m}$ is the wave number for which the wave spectrum is maximum (typically at the forcing scale)~\cite{deike2015,berhanu2018}. We keep $\epsilon\simeq0.07$ to validate the weak nonlinearity assumption from WTT.
\begin{figure}[t]
    \centering
    \includegraphics[width=1\linewidth]{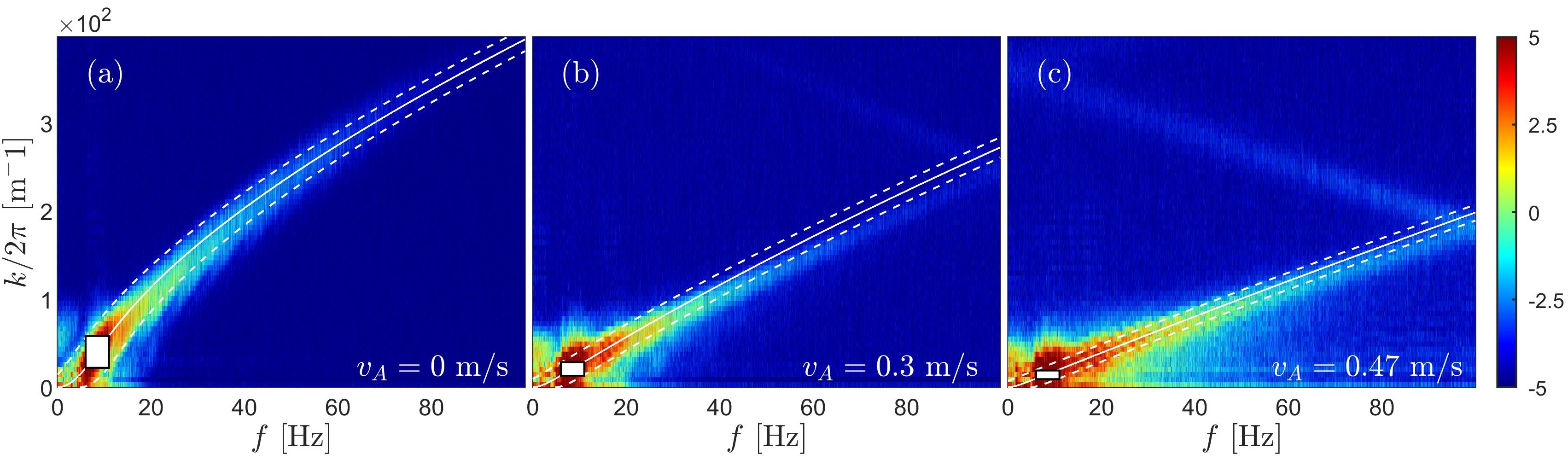} 
    \caption{Power spectrum $S_\eta(k, \omega)$ of the wave elevation for (a) $v_A=0$, (b) $v_A=0.3$, and (c)$v_A=0.47$ m/s. The constant wave steepness $\epsilon\simeq0.07$. The solid line shows the theoretical dispersion relation $\omega(k)$ of Eq.~\eqref{relat_disp} in the (a) dispersive, (b) intermediate, and (c) nondispersive cases. In the latter case, the slope of the straight line is $1/v_A$. The dashed line shows the spread dispersion relation $\omega(k)\pm\delta_\omega$ with $\delta_\omega=30$ Hz. The white rectangle shows the fixed frequency forcing range between 6 and 11 Hz. The color bar is on a logarithmic scale.}
    \label{Spatiotemp}
\end{figure}

\section{Experimental results}\label{ExpResults}
\subsection{Spatiotemporal spectral analysis}
From LSP measurements, applying to the surface elevation $\eta(x,t)$ a double space and time Fourier transform $\widehat{\eta}(k,\omega)$, we compute the spatiotemporal power spectrum $S_\eta(k,\omega)=|\widehat{\eta}(k,\omega)|^2/(\mathcal{T}L)$. Note that the signal $\eta(x,t)$ has been increased in length using its spatial symmetry to reach symmetric boundary conditions to compute $S_\eta(k,\omega)$. A Hanning windowing (hanning \textsc{Matlab} function) has also been performed to improve the quality of the spectrum. The space-time power spectra $S_\eta(k,\omega)$ are shown in Fig.~\ref{Spatiotemp} for different applied magnetic field $B$, that is, for different $v_A$. In Fig.~\ref{Spatiotemp}(a), $v_A=0$~m/s, meaning that the wave field is only driven by gravity and capillary effects. In this case, the wave energy is found to cascade over small scales and is concentrated around the gravity-capillary dispersion relation (white solid line). This is a clear indication of the presence of wave turbulence as previously reported in Ref.~\cite{Ricard2021}. A spectral broadening $\delta_\omega$ of the wave energy around this dispersion relation is also observed due to nonlinearities~\cite{Ricard2021} and is estimated\footnote{For each $k$, $\delta_\omega$ is estimated by fitting the spectrum $S_\eta(k,\omega)$ by a Gaussian function of $\omega$. The standard deviation of this fit gives an estimate of $\delta_\omega$ whose average over the $k$ values is $\delta_\omega=30$~Hz, which is almost constant for all values of $v_A$.}. When the magnetic field is increased [Figs.~\ref{Spatiotemp}(b) and \ref{Spatiotemp}(c)], the energy still cascades following the dispersion relation, but is now influenced by the magnetic effects lowering significantly the spectrum [see solid lines in Figs.~\ref{Spatiotemp}(b) and \ref{Spatiotemp}(c)]. For $v_A=0.47$~m/s, the nondispersive term in Eq.~\eqref{relat_disp} is at least ten times larger than the dispersive ones in the range of interest ($20<f<100$~Hz) as quantified in Fig.~\ref{fc_vA}. As a consequence of this quasinondispersive dispersion relation, the wave energy is then found to be concentrated around a straight line of slope close to $1/v_A$ as shown in Fig.~\ref{Spatiotemp}(c). We thus evidence a transition from a dispersive gravity-capillary wave field to a nondispersive magnetic wave field where all waves travel at a constant velocity $v_A$. The operator thus controls the dispersivity of the system via the parameter $v_A(B)$. Note that a slight mismatch between the theoretical dispersion relation and the experimental data occurs at large $v_A$. This might be due to the inhomogeneous magnetic fluctuations appearing along the wave height as explained in Sec.~\ref{MagneticEffect}. The fluctuations of the field, involving fluctuations of $v_A$, explain the mismatch but are not quantified in the present study. Note also that for $v_A=0.47$~m/s a weaker branch of the energy appears at the top of Fig.~\ref{Spatiotemp}(c). Although the maximum visible frequency in the spectrum is $f_e/2=100$~Hz, i.e., half the sampling frequency, energy at higher frequencies $f>f_e$, i.e., $k/(2\pi)>198$ m$^{-1}$ for $v_A=0.47$ m/s, can be seen, however, due to the spectrum aliasing effect. Despite viscous effects acting from about 100~Hz, energy occurring at higher frequencies is a consequence of singularities that give energy to all frequencies (see below). Note also that no other coherent structure such as bound waves appears in Fig.~\ref{Spatiotemp}. 

\subsection{Surface elevation signals and time-frequency analysis}\label{signal}
\begin{figure}[h!]
    \centering
    \includegraphics[width=0.72\linewidth]{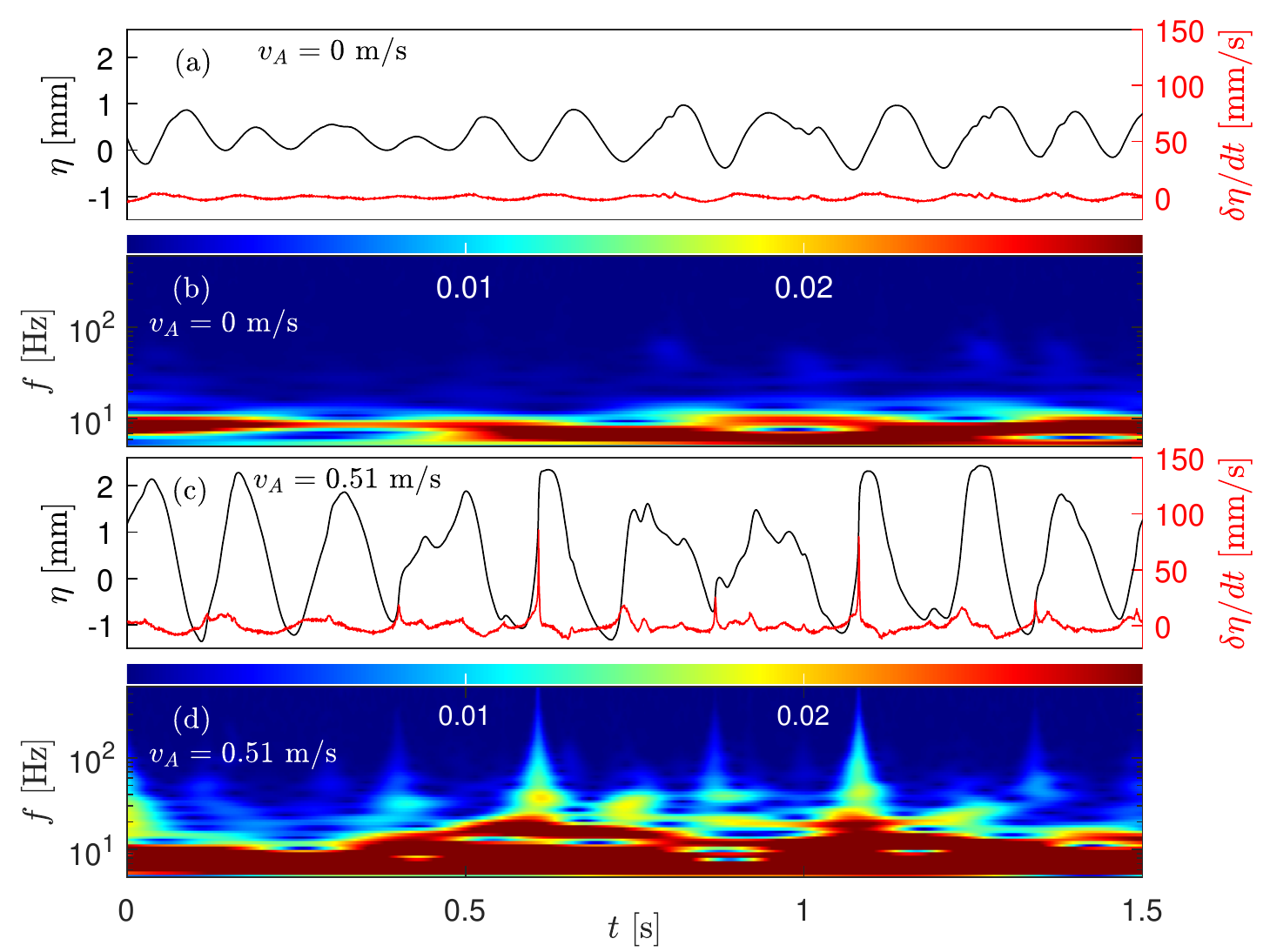} 
    \caption{(a) Typical temporal evolution of the surface elevation $\eta(t)$ (black line) and its first-order difference $\delta\eta(t)/dt$ (red line) and (b) corresponding time-frequency spectrum of $\eta(t)$ obtained by a wavelet transform, for the dispersive case ($v_A=0$ m/s). (c) and (d) Same as in (a) and (b) but for the nondispersive case ($v_A=0.51$ m/s). Here $\epsilon\simeq0.07$ in the two cases.}
    \label{wavelet_signal_0}
\end{figure}

Typical temporal signals of the surface elevation $\eta(t)$ (black line) and of its first-order difference $\delta\eta(t)=\eta(t+dt)-\eta(t)$ (red lines) are shown in Fig.~\ref{wavelet_signal_0}(a) for the dispersive case ($v_A=0$~m/s) and in Fig.~\ref{wavelet_signal_0}(c) for the nondispersive case ($v_A=0.51$~m/s). We also compute the corresponding wavelet transforms (using the continuous 1D wavelet transform \textsc{Matlab} function)~\cite{grossmann1984} to obtain a time-frequency analysis of the energy spectra as plotted in Figs.~\ref{wavelet_signal_0}(b)--\ref{wavelet_signal_0}(d) (see Appendix~\ref{2spectra} for longer signals). The wavelet transform is preferred to a short-time Fourier transform, e.g., spectrogram, that has issues with the frequency-time resolution trade-off. For the dispersive case ($v_A=0$~m/s), no coherent structure appears for the temporal evolution of the surface elevation, its first-order difference $\delta\eta$ remaining close to 0. For the nondispersive case ($v_A=0.51$~m/s) the typical wave height is found to increase, whereas some peaks occur in its first-order difference corresponding to discontinuities in $\eta(t)$. As discussed in Sec.~\ref{MagneticEffect}, a concentration of the magnetic field lines occurs at the crests and troughs of the wavy interface to satisfy the magnetic boundary conditions at the interface~\cite{RosensweigBook}, leading to a stronger magnetic field and so to a stronger value of $v_A$ at the wave crest than at its base. Thus, for a given wave, $v_A$ depends on the vertical coordinate $z$ with $\partial v_A/\partial z>0$. Since the wave crest is faster than its base, it ends up creating a discontinuity, i.e., a singularity, called afterward shock wave. Shock waves are also visible in the wavelet spectrum [Fig.~\ref{wavelet_signal_0}(d)], where energy is present at all frequencies even beyond the viscous scale of the order of 100 Hz. Although subjected to dissipation during their propagation, shock waves are thus coherent structures rich in the frequency domain. Note that the Maxwell stress which should decrease the wave height in the magnetic field direction~\cite{RosensweigBook,Kochurin2022} is not reported here. This higher-order effect could occur at higher $v_A$ not experimentally achievable in our parameter range (see Appendix~\ref{stdva}). 

\begin{figure}[h!]
    \centering
    \includegraphics[width=0.5\linewidth]{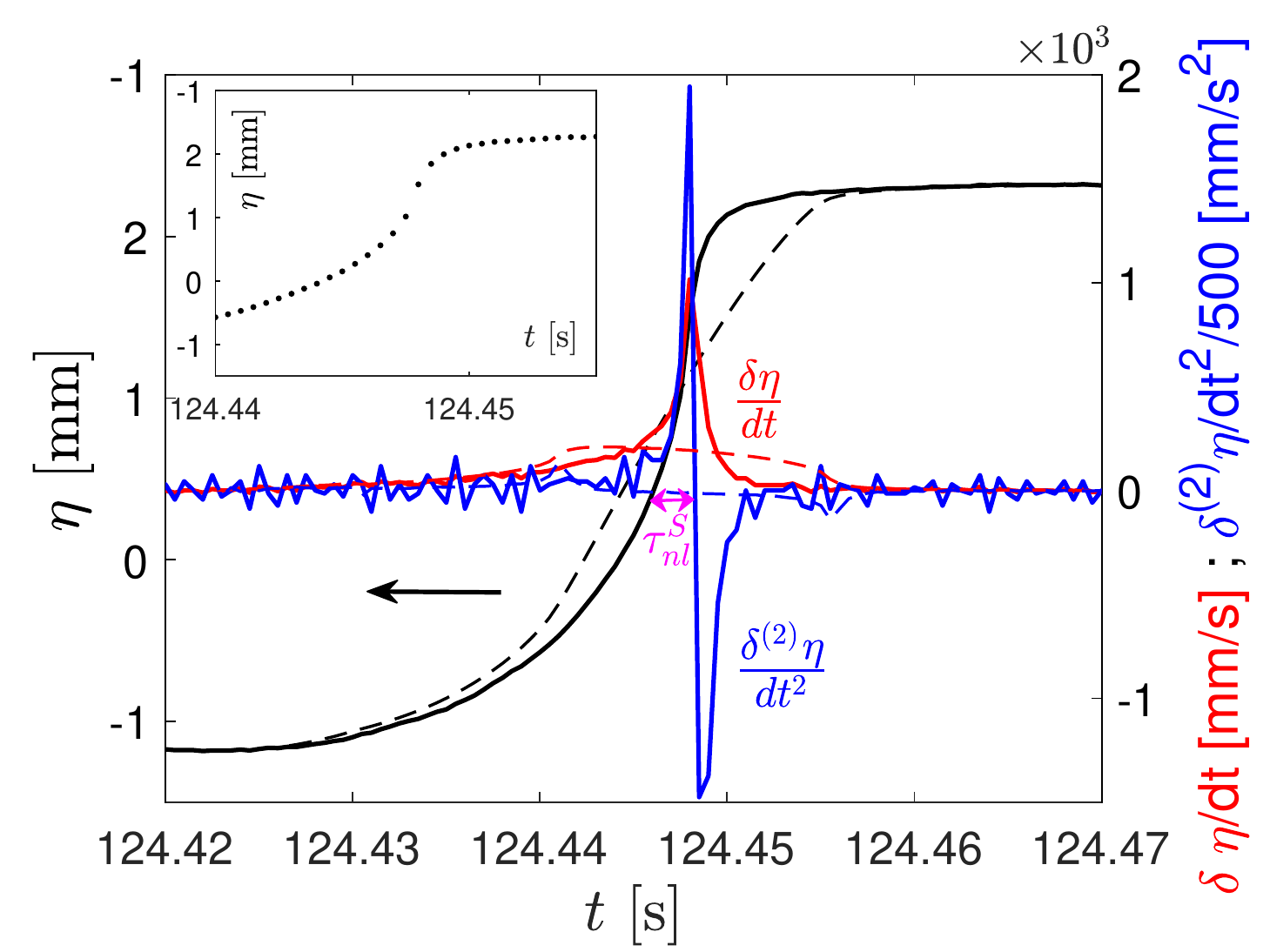} 
    \caption{Enlargement of a typical shock wave $\eta(t)$ (black solid line), its first-order difference $\delta\eta(t)/dt$ (red solid line) and its second-order one $\delta^2\eta(t)/dt^2$ (blue solid line) in the nondispersive case ($v_A=0.51$ m/s). The value of $\delta^{(2)}\eta/dt^2$ is divided by 500 to observe it on the same vertical scale as $\delta\eta/dt$. Dashed lines show the same but when the singularity is removed by numerical postprocessing, thus smoothing the signals. The black arrow shows the direction of wave-front propagation.  The purple arrow shows the nonlinear timescale of a shock wave, $\tau_{nl}^S$, (see Sec.~\ref{Sectimescale}). The inset shows the enlargement of the shock wave with only experimental discrete data to evidence the jump at the second-order discontinuity.}
    \label{Shock}
\end{figure}

A typical shock wave signal $\eta(t)$ and its first- and second-order differences $\delta\eta(t)=\eta(t+dt)-\eta(t)$ and $\delta^{(2)}\eta(t)=\eta(t+2dt)-2\eta(t+dt)+\eta(t)$ respectively, are plotted in Fig.~\ref{Shock} for the nondispersive case ($v_A=0.51$~m/s). We checked that this localized singularity keeps its shape and travels along the canal at constant velocity with no breaking (see Appendix~\ref{1shock} for the displacement of a single shock along the canal). The discontinuity in $\eta(t)$ displayed in Fig.~\ref{Shock} corresponds to a rather long peak in its first-order difference $\delta\eta$ and to a very thin peak in its second-order difference $\delta^{(2)}\eta$. This short peak is assumed to be close to a Dirac peak, to claim that the singularity observed here is of second order. It is worth noting that the nondispersive shock waves observed here do not exhibit a fully vertical front. This observation is emphasized in the inset of Fig.~\ref{Shock}, where only the experimental discrete data of the shock wave are plotted. A jump in the signal is visible corresponding to a second-order discontinuity of $\eta(t)$. Although a fully vertical shock cannot be measured with a single-point gauge, the spatiotemporal measurement of the shock-wave shape confirms that the latter does not reach a fully vertical front (see Appendix~\ref{1shock}). The shocks observed therefore differ from classical shock waves driven by the 1D Burgers equation displaying singularities of the first order (Dirac-$\delta$ distribution in their first-order difference)~\cite{burgers1948}. Even if its amount is small, dispersive effects might prevent the formation of a vertical-front shock wave, and it is difficult to say if higher $v_A$ values would lead to a vertical front since the Maxwell stress would occur, flattening the waves. Note that each singularity in the system can be removed by numerical postprocessing, leading, as expected, to smoothing the signal around the discontinuity (see dashed lines in Fig.~\ref{Shock}). 

\begin{figure}[t!]
    \centering
    \includegraphics[width=0.8\linewidth]{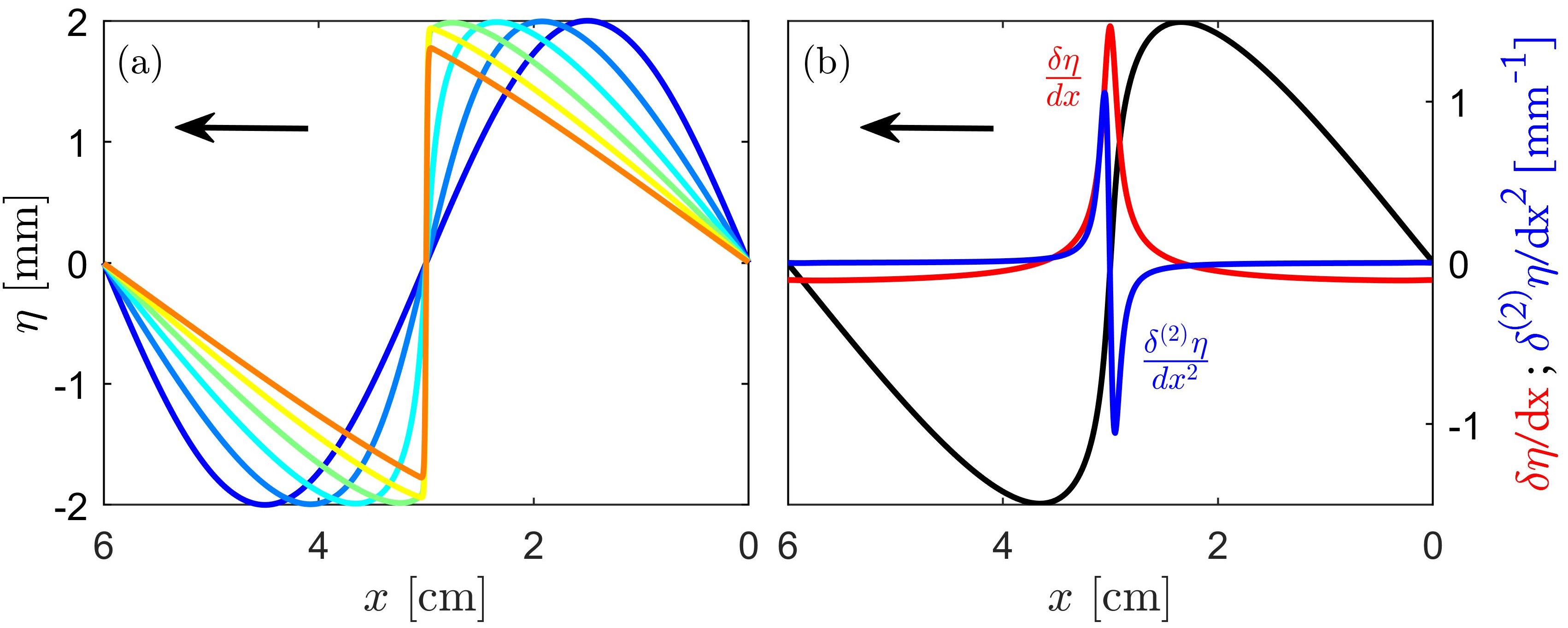} 
    \caption{Numerical solution of the 1D Burgers equation, $\eta(x,t)$, following an implicit scheme from a sinusoidal initial condition at $t=0$ (blue). (a) Solutions for increasing values of $t$ (from blue to orange). (b) Solution at a fixed time $t$ (just before reaching the vertical front) along with the corresponding first- ($\delta\eta/dx$, red solid line) and second-order difference ($\delta^{(2)}\eta/dx^2$, blue solid line). The abscissa is from right to left to be consistent with the experimental temporal measurements in Fig.~\ref{Shock}. The arrows show the direction of the wave-front propagation.}
    \label{BurgerSimmu}
\end{figure}

To compare the typical shape of our coherent structures (Fig.~\ref{Shock}), we solve numerically the 1D Burgers equation~\cite{burgers1948}
\begin{equation}
    \frac{\partial\eta}{\partial t}+A\eta\frac{\partial\eta}{\partial x}=\nu\frac{\partial^{2}\eta}{\partial x^2},
\end{equation}
with $A=v_A/d$ ($v_A=0.5$~m/s and $d=2$~cm) a constant chosen for dimensional homogeneity and $\nu=2.86\times10^{-6}$~m$^2$/s the kinematic viscosity of the liquid. We use an implicit scheme using the Crank-Nicolson formulation~\cite{crank1947} and a Thomas algorithm~\cite{Thomas49} with the initial condition $\eta(x,t=0)=\sin(x)$. The numerical grid is resolved with 1024 points. The results are plotted for different times in Fig.~\ref{BurgerSimmu}(a). As expected, a steepening of the wavefront appears before dissipation decreases slightly the amplitude of the shock. This kind of vertical shock would lead to breaking experimentally. No fully vertical front appears experimentally, but rather a shape close to the one obtained just before the Burgers shock, and that is conserved over time [see Figs.~\ref{Shock} and~\ref{BurgerSimmu}(b)]. Moreover, this shock-wave shape exhibits both for the numerical and experimental results a long peak on the first-order difference $\delta\eta$ and a short peak (similar to a Dirac one) on the second-order difference $\delta^{(2)}\eta$. To sum up, as a consequence of the magnetic effects, this nondispersive system generates coherent structures that are close to the Burgers shock waves with a slightly less steep front (less than 0.1\%) and a self-similar shape that is conserved over time. It is worth noting that even if strong similarities occur between the numerical results of the Burgers equation and the experimental results found here, e.g., the presence of shock waves and nondispersive system, no rigorous analytical link is established in the present study. The link is only qualitative (see Sec.~\ref{PowerSpectrum} for power spectra and  Sec.~\ref{ShockStats} for probability density functions of the surface elevation) but provides some interesting insights that deserve further theoretical work.

\subsection{Experimental wave energy spectra}\label{PowerSpectrum}
\begin{figure}[h!]
    \centering
    \includegraphics[width=0.49\linewidth]{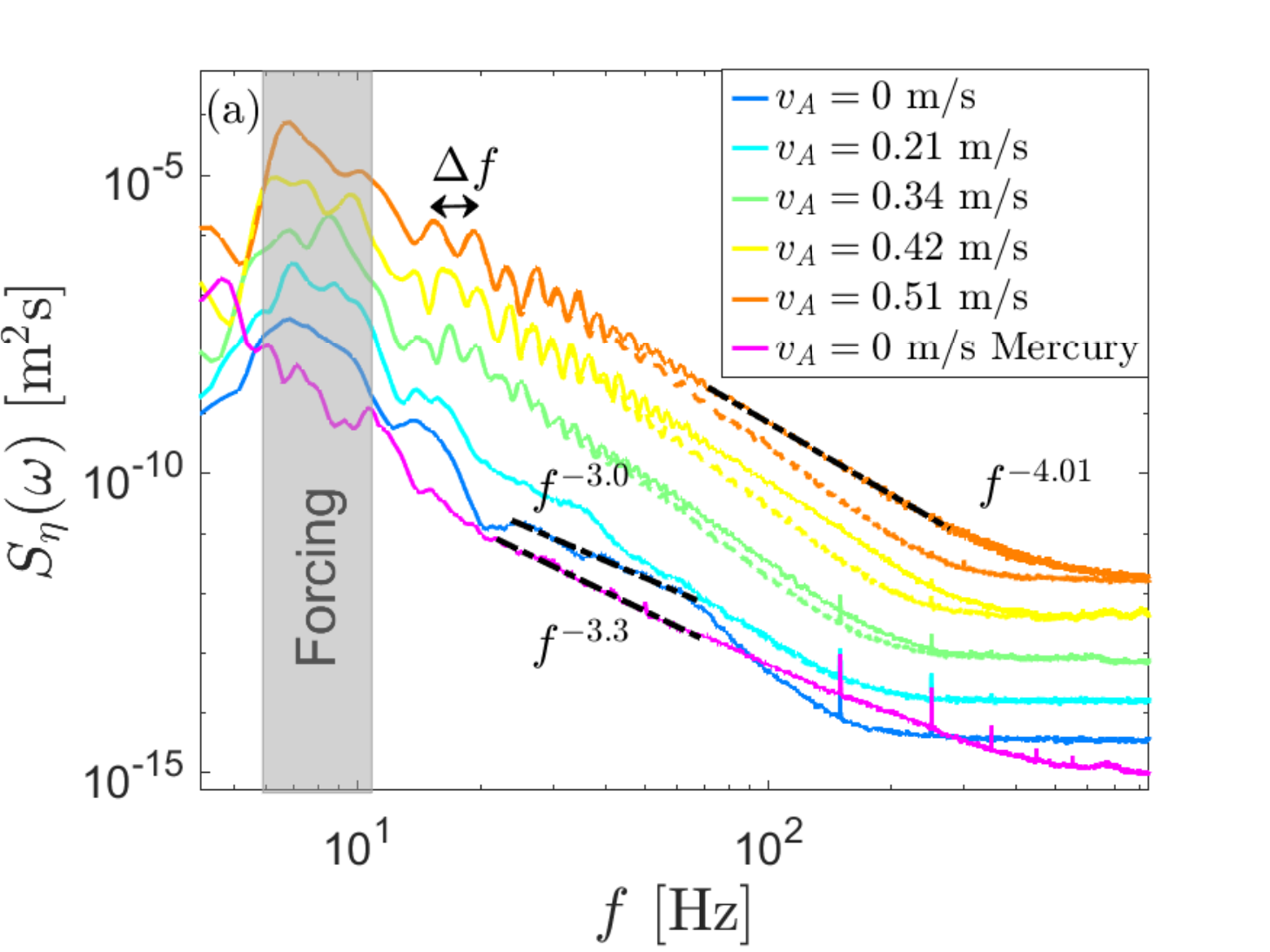} 
    \hfill
    \includegraphics[width=0.49\linewidth]{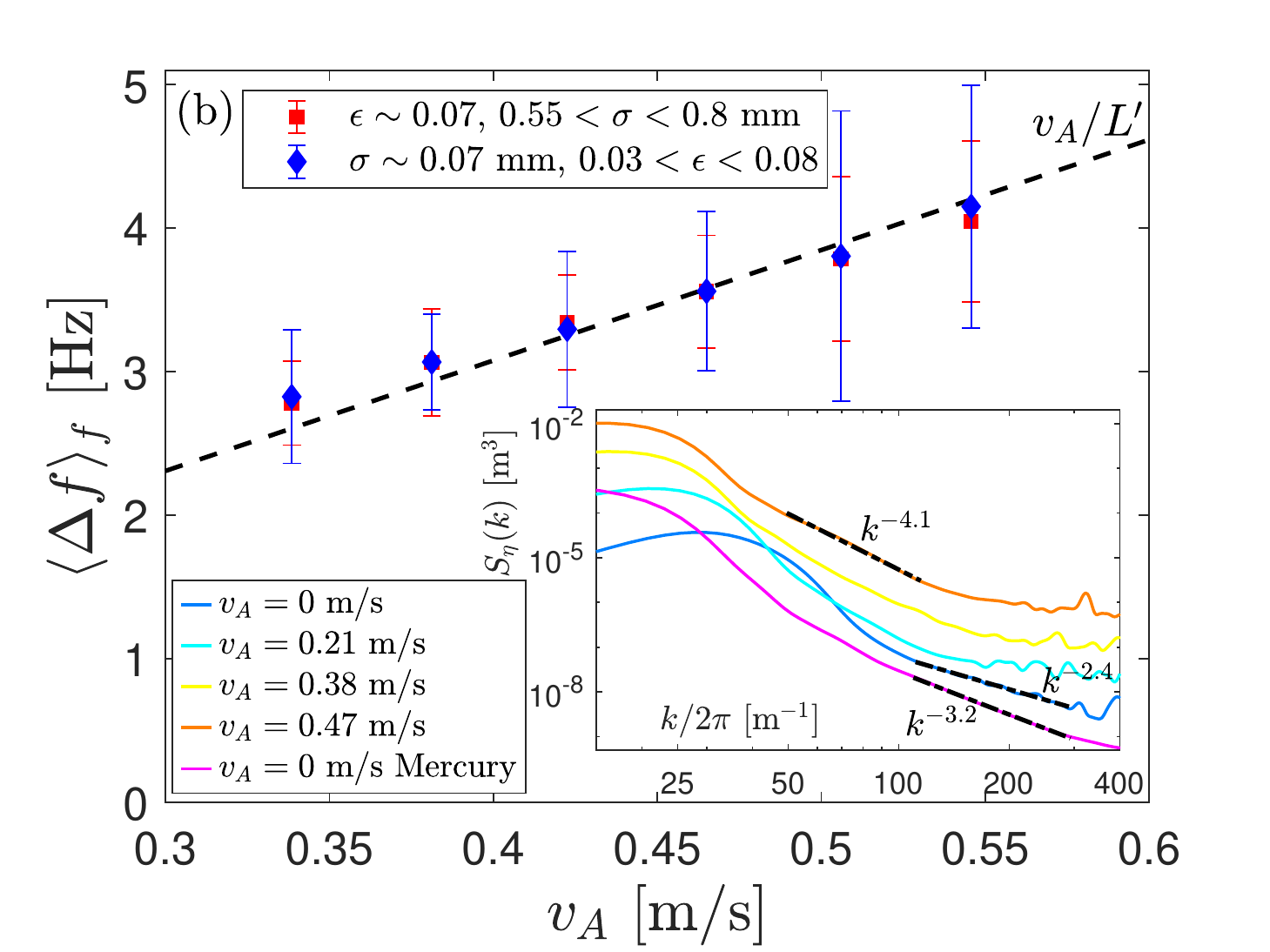} 
    \caption{(a) Frequency spectra $S_\eta(f)$ for different $v_A$ (solid lines) and $\epsilon \simeq 0.07$ on a log-log plot. Spectra have been shifted vertically for clarity. The dashed lines show the same but with the singularities removed from the signal. The gray area is the frequency bandwidth of the random forcing. The black dash-dotted line shows $f^{-4.01}$ best fit for $v_A=0.51$ m/s, $f^{-3.0}$ best fit for $v_A=0$ m/s, and $f^{-3.3}$ best fit for $v_A=0$ m/s using mercury~\cite{Ricard2021}. Here $\Delta f$ is the frequency difference occurring between two successive spectrum peaks. (b) Evolution of the mean frequency gap $\langle\Delta f\rangle_f$ between local spectral peaks as a function of $v_A$ for two sets of forcing, either at constant $\epsilon$ or at constant standard deviation $\sigma$ of the surface elevation. The dashed line shows the best linear fit in $v_A/L'$ with $L'=13$ cm, the available canal length. Error bars come from the standard deviation of the measurement of $\Delta f$. The inset shows wave-number power spectra $S_\eta(k)$ for different $v_A$ and $\epsilon \simeq 0.07$ on a log-log plot. Spectra have been shifted vertically for clarity. Black dash-dotted lines show the best fits in $k^{-4.1}$ for $v_A=0.47$ m/s, $k^{-2.4}$ for $v_A=0$ m/s using ferrofluid, and $k^{-3.2}$ for $v_A=0$ m/s using mercury~\cite{Ricard2021}.}
    \label{Spectromeg}
\end{figure}

The frequency power spectrum $S_\eta(\omega)\equiv |\widehat{\eta}(\omega)|^2/\mathcal{T}$ is now computed from the single-point measurement of the surface elevation $\eta(t)$ using its temporal Fourier transform $\widehat{\eta}(\omega)$. $S_\eta(\omega)$ is shown in Fig.~\ref{Spectromeg}(a) for different dispersion strengths, i.e., different $v_A$. For the dispersive case ($v_A=0$~m/s), the wave spectrum follows a power-law cascade characteristic of wave turbulence although occurring over a rather small inertial range (bottom blue curve). This frequency range (between 20 and 70~Hz) corresponds to the entanglement of gravity and capillary effects, whereas no pure capillary wave turbulence is observed here due to viscous effects ($f \gtrsim 70$~Hz). Note that the exponent of this frequency power law $S_\eta(\omega)\sim \omega^{-3.0\pm0.3}$ is close to what was obtained with a low-viscosity fluid, e.g., mercury with $S_\eta(\omega)\sim \omega^{-3.3\pm0.3}$, within a similar gravity-capillary frequency range [see purple curve in Fig.~\ref{Spectromeg}(a)]~\cite{Ricard2021}.

For quasi-nondispersive cases (high enough $v_A$), two phenomena are visible on the power spectra. The first one is the emergence of well-defined series of local peaks. These peaks are found to be separated by a frequency gap $\Delta f$ which is nearly constant for a single value of $v_A$. The frequency gap is averaged for each spectrum and plotted against $v_A$ in Fig.~\ref{Spectromeg}(b). Two sets of measurements corresponding to two different forcing are plotted and are well fitted linearly by $\langle\Delta f\rangle_f=v_A/L'$, i.e., $\langle\Delta \omega\rangle_f= v_A(2\pi/L')$ with $L'=13$~cm the length of the canal $L$ minus the gap filled by the wave maker ($\sim 2$~cm). As all waves travel with the same nondispersive velocity, they are then detected by the single-point gauge every same time $1/\Delta f$. This implies the emergence of peaks of frequencies that are directly linked to the main eigenmode of the canal $2\pi/L'$. Finite-size effects thus emerge experimentally because of the nondispersivity.

The second effect of the nondispersivity is visible at high frequencies of the power spectra. A very-well-defined power law appears on one decade in the range $f\in[30,300]$ Hz, thus well beyond the beginning of viscous effects around 100~Hz. This cascade scales in $S_\eta(\omega)\sim\omega^{-4.01\pm0.05}$ and is found to agree with the Kuznetsov spectrum of singularities of second order $n=2$, i.e., Dirac-$\delta$ distribution on the second-order difference $\delta^{(2)}\eta$, conserving their shape, i.e., $\omega\sim k$~\cite{KuznetsovJETP2004}. The shock waves present in the signal thus spread energy at all frequency scales. When removing the singularities from the signal (as in Fig.~\ref{Shock}), the previous well-defined power law in $\omega^{-4}$ in the power spectrum disappears and dissipative effects seem to drive the cascade after 90 Hz [see dashed lines in Fig.~\ref{Spectromeg}(a)]. These results evidence a transition from gravity-capillary wave turbulence, in the dispersive case ($v_A=0$~m/s), for which the cascade mechanism is due to resonant interactions between weakly nonlinear waves, to a nondispersive regime ($v_A=0.51$~m/s) where the energy is mainly concentrated in second-order singularities ($n=2$) and dissipated by viscous effects. 

Using the spatiotemporal measurements averaged over time, the wave-number power spectrum $S_\eta(k)$ is plotted in the inset of Fig.~\ref{Spectromeg}(b) for different values of $v_A$. At $v_A=0$~m/s, a power law in $k^{-2.4\pm0.1}$ is observed due to gravity-capillary wave turbulence. This power-law exponent differs from the one found for a much less viscous fluid, i.e., mercury~\cite{Ricard2021}, plotted also in the inset of Fig.~\ref{Spectromeg}(b). At large $v_A$, a steeper power law in $k^{-4.1\pm0.1}$ is found, and the exponent is close to the one found for the frequency power spectrum $S_\eta(\omega)\sim\omega^{-4.01}$. This similarity thus confirms that a nondispersive regime is achieved since the two spectra are linked by $S_\eta(k)dk=S_\eta(\omega)d\omega$ using $\omega \sim k$. The spectrum close to $k^{-4}$ is hence a spectrum of second-order discontinuities due to shock waves, which supports the conclusion made with the temporal spectrum on the second-order singularities. Note that, in the inset of Fig.~\ref{Spectromeg}(b), the forcing scale moves to smaller $k$ with increasing $v_A$ as a result of Eq.~\eqref{relat_disp} with a constant forcing frequency. Note also that because of the lowering of the dispersion relation with increasing $v_A$ as observed in Fig.~\ref{Spatiotemp}, the measurement noise level appears at $k/2\pi>400$~m$^{-1}$ for $v_A=0$~m/s and at $k/2\pi>200$~m$^{-1}$ for $v_A=0.47$~m/s. The statistics of the shock waves will be thus performed in Sec.~\ref{ShockStats} using the single-point measurements due to their better resolution and signal-to-noise ratio than the spatiotemporal ones.

\subsection{Energy flux} \label{SecFlux}
Weak turbulence theory aims to describe wave turbulence but requires strong hypotheses~\cite{ZakharovBook,NazarenkoBook,Newell2011}. In particular, WTT assumes a constant energy flux during the energy cascade through the scales. In this section, we test this hypothesis when the wave turbulence regime occurs ($v_A=0$~m/s) and how the energy flux departs from a constant when reaching the shock-wave regime (at high $v_A$).

\begin{figure}[t!]
    \centering
    \includegraphics[width=0.5\linewidth]{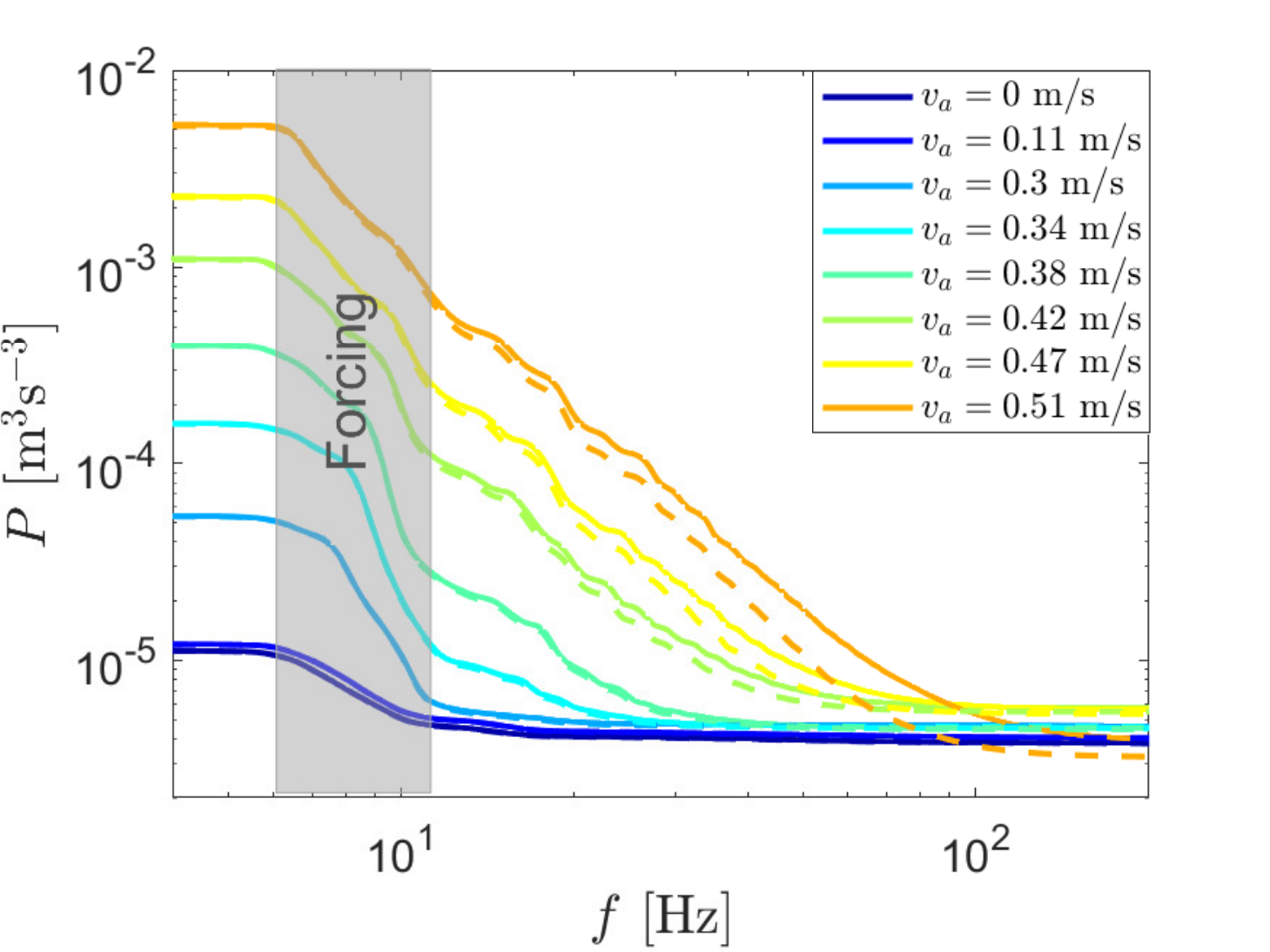} 
       \caption{ Evolution of the indirectly measured energy flux $P$ with the frequency $f$, for $\epsilon\simeq0.07$. Solid lines correspond to different $v_A$. Dashed lines show the same but with singularities removed by signal postprocessing. The gray area indicates the forcing frequency bandwidth.}
    \label{Flux_time}
\end{figure}

The energy flux $P$ is computed as $P(\omega^*)=\int_{\omega^*}^{\omega_{m}}E(\omega)D(\omega)d\omega$ with $E(\omega)=gS_\eta(\omega)+v_A^2kS_\eta(\omega)+(\gamma/\rho)k^2S_\eta(\omega)$ the total wave-energy density, $D=k(\omega)\sqrt{\nu\omega/2}$ the main contribution of the viscous energy dissipation rate for a contaminated interface~\cite{Ricard2021,berhanu2018,Lamb1932,deikePRE2014}, $\omega_{m}/(2\pi)=1000$ Hz, and $k(\omega)$ as in Eq.~\eqref{relat_disp}. The variation of $P$ over frequency scales is plotted in Fig.~\ref{Flux_time} for different $v_A$ with (solid lines) and without  (dashed lines) shock waves. For low values of $v_A$, the energy flux is, as expected, constant in the inertial range (as in~\cite{Ricard2021}), showing that no dissipation occurs in this range and that the energy cascades over scales continuously because of wave interactions following WTT predictions. $P$ is found to increase with $v_A$ as a consequence of the increase of the energy at the forcing scales that is required to keep a constant wave steepness [because $v_A$ increases the wavelength as shown by Eq.~\eqref{relat_disp}]. Furthermore, for large values of $v_A$, $P$ is no longer constant and is found to decrease with $f$. This can be explained by dissipation that occurs at all scales~\cite{deikePRE2014}. As shock waves travel by conserving their shape, they transport energy over space without any interactions. While they transport this energy, viscous dissipation occurs reducing their amplitude until they disappear (see Appendix~\ref{1shock}). Note that the presence of the discontinuity in the shock wave does not have any significant impact on the energy flux (solid and dashed lines are almost superimposed in Fig.~\ref{Flux_time}), even when the discontinuity is removed, the energy is still in the shock wave and continues to travel and to be dissipated.

\subsection{Timescales}\label{Sectimescale}
\begin{figure}[b!]
    \centering
    \includegraphics[width=0.5\linewidth]{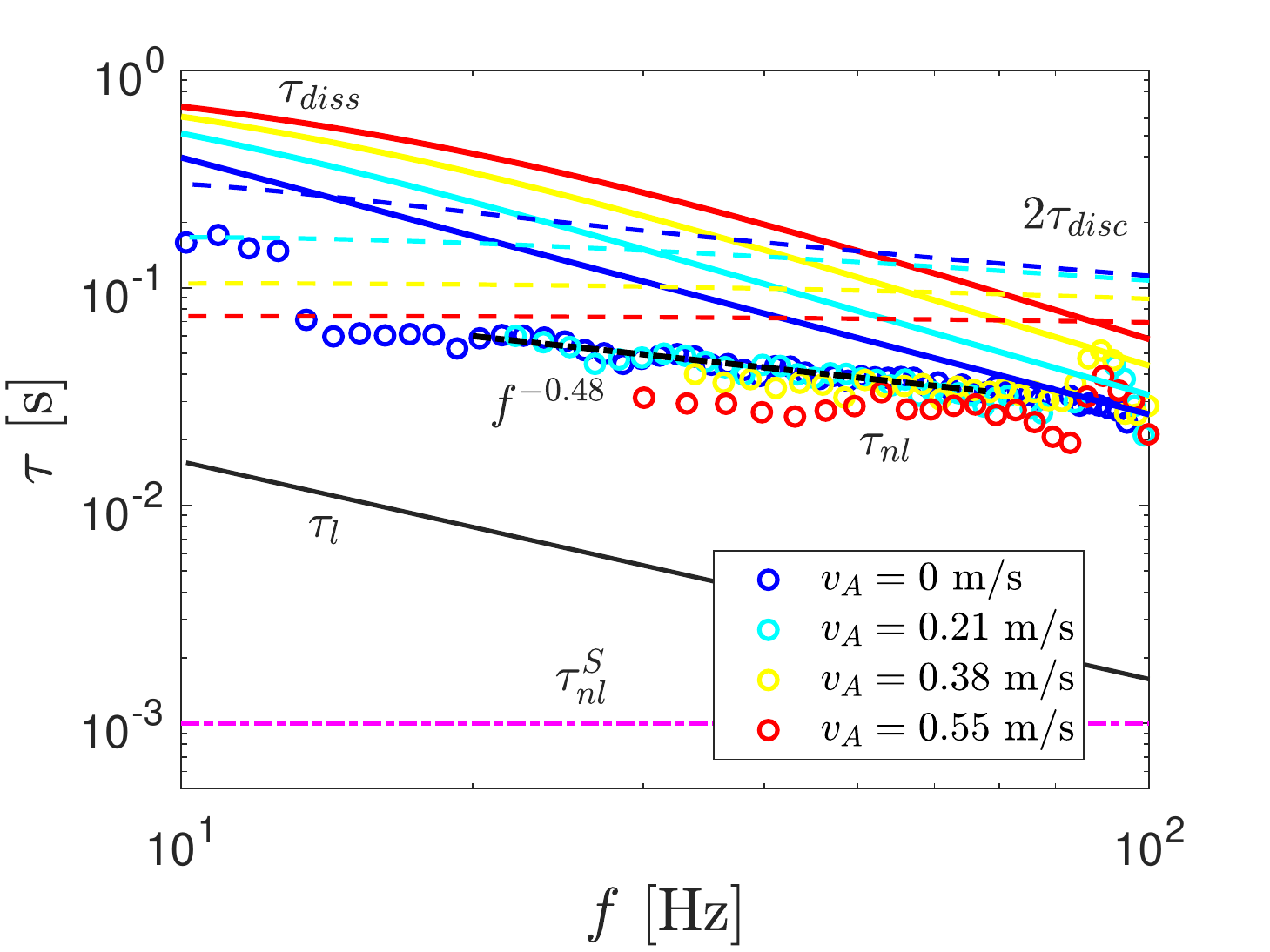} 
    \caption{Wave turbulence timescales as a function of the frequency scale $f$ for different $v_A$. The solid black line shows the linear timescale $\tau_{l}=1/\omega$. Circles show the nonlinear timescale $\tau_{\mathrm{nl}}$ estimated from Fig.~\ref{Spatiotemp}. Colored solid lines show the linear viscous dissipation timescale $\tau_{\mathrm{diss}}$ (see the text). Colored dashed lines show the discreteness time $\tau_{\mathrm{disc}}$ (see the text). The purple dash-dotted line shows the nonlinear shock-wave timescale $\tau_{\mathrm{nl}}^S$ estimated from Fig.~\ref{Shock}.}
    \label{TimescaleFig}
\end{figure}
We now test another WTT assumption, namely, the timescale separation between the linear time $\tau_l$, the nonlinear time $\tau_{\mathrm{nl}}$, the dissipation time $\tau_{\mathrm{diss}}$ (quantifying dissipative effects), and the discreteness time $\tau_{\mathrm{disc}}$ (quantifying finite-size effects of the canal)~\cite{ARFM2022,CazaubielPRL2019,Ricard2021}. Indeed, WTT assumes~\cite{NazarenkoBook}
\begin{equation}
    \tau_l(\omega)\ll\tau_{\mathrm{nl}}(\omega)\ll[\tau_{\mathrm{diss}}(\omega); \tau_{\mathrm{disc}}(\omega)], 
    \label{sclae_sep}
\end{equation}
regardless of $\omega=2\pi f$ in the inertial range. The nonlinear evolution is thus assumed to be slow compared to the fast linear oscillations (wave period) but short compared to the typical wave dissipation time and the time linked to finite-size effects, enabling then an energy cascade to occur in the inertial range. The evolutions of these timescales with $f$ are plotted in Fig.~\ref{TimescaleFig}. The linear timescale is defined as $\tau_l=1/\omega$ (black solid line). The nonlinear timescale $\tau_{\mathrm{nl}}$ (colored circles) is estimated by the broadening of the energy around the dispersion relation as $1/\delta_\omega$ (see Fig.~\ref{Spatiotemp}). $\tau_{\mathrm{nl}}$ follows a frequency power law close to $f^{-1/2}$ and decreases slightly with $v_A$. The dissipation timescale $\tau_{\mathrm{diss}}$ (colored solid lines) is computed as $\tau_{\mathrm{diss}}=2\sqrt{2}/[k(\omega)\sqrt{\nu\omega}]$, the main viscous contribution from the surface boundary layer with an inextensible film~\cite{Lamb1932,deikePRE2014}. This time increases with $v_A$ meaning that dissipative effects are less significant at high $v_A$. This effect can be observed in the spectra in Fig.~\ref{Spectromeg}(a), even when the discontinuities are removed (energy is present until $250$~Hz for $v_A=0.51$~m/s and less than $150$~Hz for $v_A=0$~m/s). The discreteness time $\tau_{\mathrm{disc}}$ (colored dashed lines) is computed as $\tau_{\mathrm{disc}}=1/\Delta \omega_{\mathrm{disc}}$ with $\Delta \omega_{\mathrm{disc}}=(\partial\omega/\partial k)\Delta k$ and $\Delta k =2\pi/L'$ the first eigenmode of the canal~\cite{ARFM2022}. No discreteness effect is expected for $\tau_{\mathrm{nl}}(\omega)<2\tau_{\mathrm{disc}}(\omega)$, i.e., when the nonlinear spectral widening is larger that the half-frequency separation between adjacent eigenmodes. This discreteness time decreases with increasing $v_A$, meaning that finite-size effects are more significant at large $v_A$. These effects are highlighted in the spectra of Fig.~\ref{Spectromeg}(a) by the emergence of well-defined series of local peaks separated by a constant frequency gap $\Delta \omega=v_A (2\pi/L')$ [see Fig.~\ref{Spectromeg}(b) and Sec.~\ref{PowerSpectrum}]. Note that, neglecting gravity and capillary effects, $\Delta \omega=\Delta \omega_{\mathrm{disc}}=1/\tau_{\mathrm{disc}}$. Figure~\ref{TimescaleFig} then evidences that the timescale separation of Eq.~\eqref{sclae_sep} is well validated experimentally in the inertial range, for all values of $v_A$. However, it is worth noting that the estimation of $\tau_{\mathrm{nl}}$ from the spatiotemporal spectrum of Fig.~\ref{Spatiotemp} does not include shock waves (as they do not explicitly appear in such a plot). To solve this issue, we define another nonlinear timescale $\tau_{\mathrm{nl}}^S$ that only takes into account the shock waves (purple dash-dotted line). $\tau_{\mathrm{nl}}^S$ is defined as the width of the corresponding peak of the second-order difference $\delta^{(2)}\eta/dt^2$ (see Fig.~\ref{Shock}). We find $\tau_{\mathrm{nl}}^S\sim10^{-3}$~s which is of the same order of magnitude for every shock wave regardless of the value of $v_A$. Figure~\ref{TimescaleFig} then shows that $\tau_l(\omega) > \tau_{\mathrm{nl}}^S(\omega)$ which means that when shock waves are prevalent the timescale separation hypothesis is no longer verified and a critical balance is achieved~\cite{NazarenkoBook}. This supports the fact that the energy is stored in coherent structures at large enough $v_A$, whereas at low $v_A$ an energy transfer through the scales occurs due to wave turbulence.

\begin{figure}[t]
    \centering
    \includegraphics[width=0.49\linewidth]{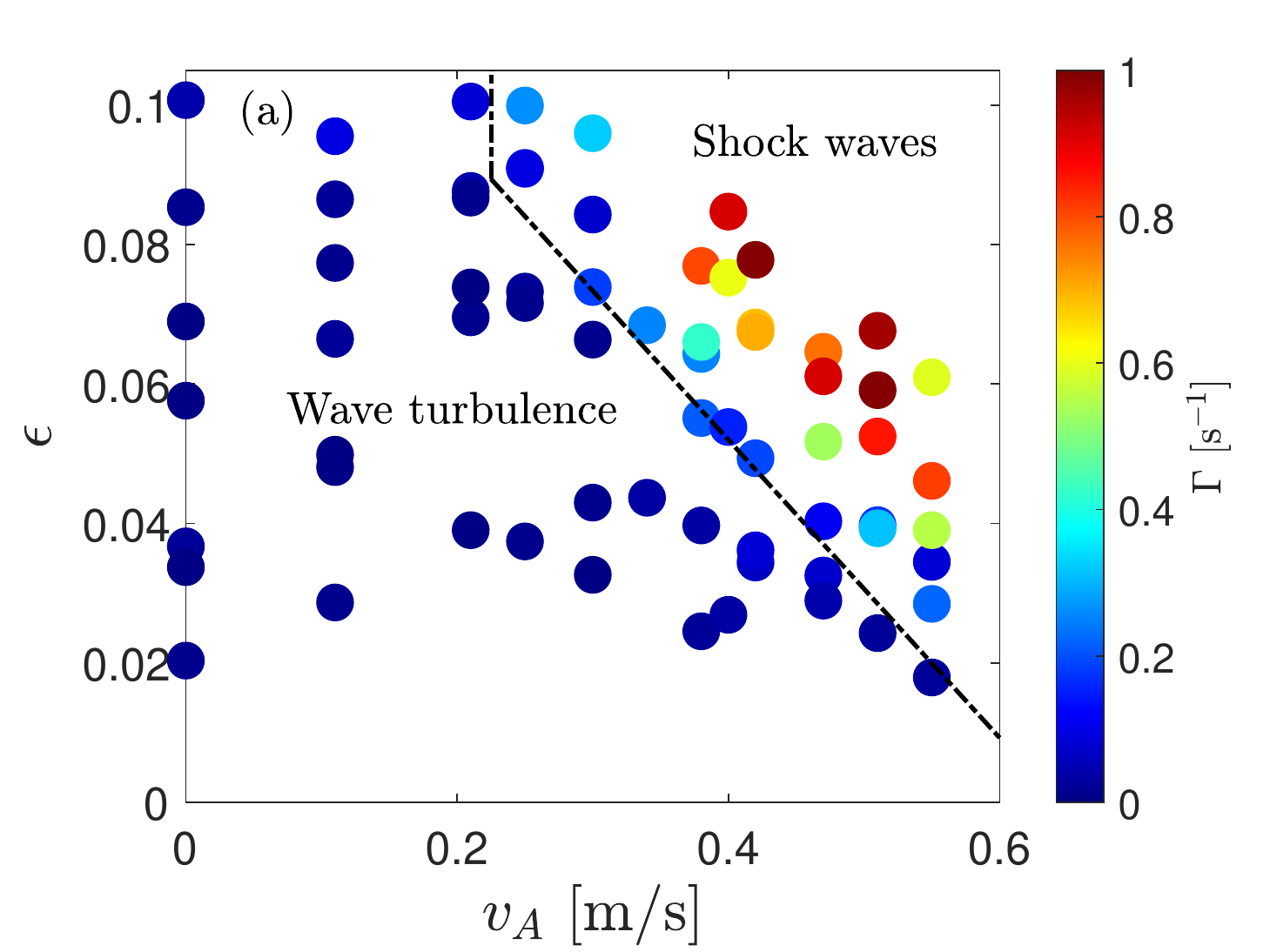} 
    \hfill
    \includegraphics[width=0.49\linewidth]{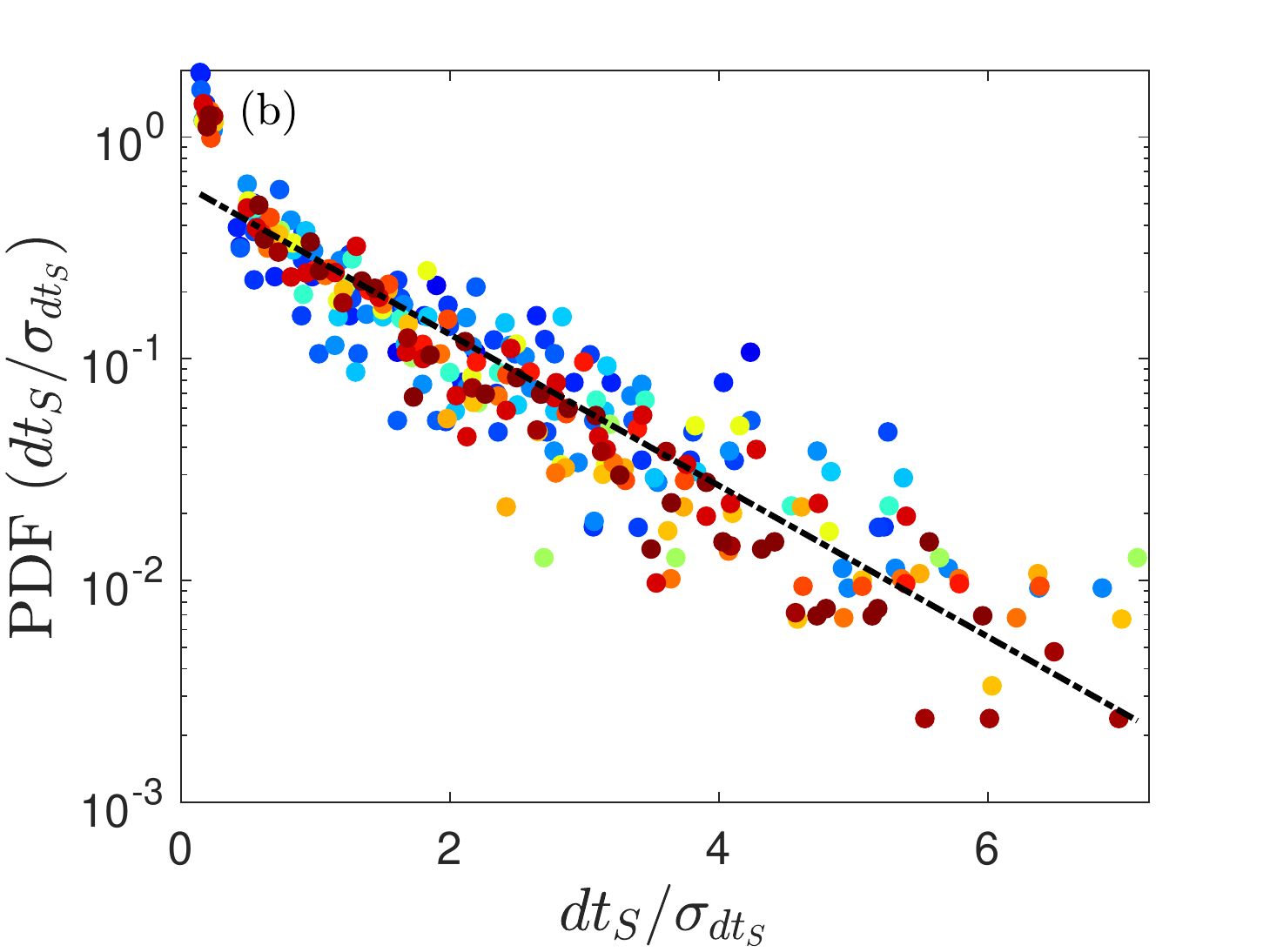} 
    \caption{(a) Phase diagram between the dispersive wave turbulence regime and the shock-wave regime as a function of the magnetic parameter $v_A$ and the wave steepness $\epsilon$. The dash-dotted line distinguishes the predominance of each regime (random waves or localized shock-wave structures) and corresponds to a fixed shock rate $\Gamma\approx0.1$ s$^{-1}$. (b) PDF of the time $dt_S$ between two successive shocks for all $v_A$ values and $\Gamma>0.1$~s$^{-1}$. The color bar is the same as in (a). The black-dashed line is the best fit in $e^{-0.78 dt_S}$.}
    \label{DiagPhase}
\end{figure}

\begin{figure}[t]
    \centering
    \includegraphics[width=0.96\linewidth]{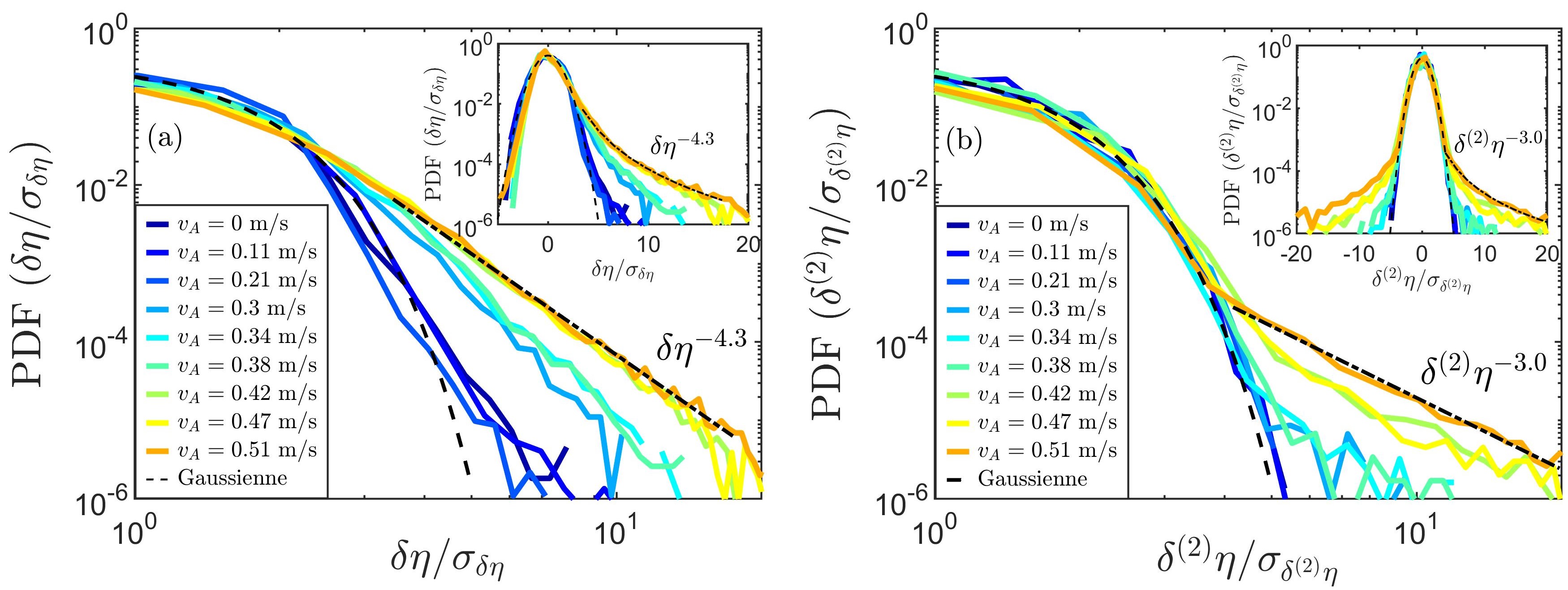}  
    \caption{(a) Probability distribution functions of the first-order difference $\delta\eta$ normalized by its standard deviation $\sigma_{\delta\eta}$ for increasing $v_A$ (from blue to orange) for $\epsilon\simeq0.07$ on a log-log scales. The dashed line shows the Gaussian distribution. The dash-dotted line shows the best power-law fit in $\delta\eta^{-4.3}$ for $v_A=0.51$~m/s. The inset shows the same on a semilogarithmic scale. (b) Same as in (a) but for the second-order difference $\delta^{(2)}\eta$ with the best power-law fit in $\delta^{(2)}\eta^{-3.0}$ for $v_A=0.51$~m/s.}
    \label{pdf_grad}
\end{figure}

\section{Shock-wave statistics}\label{ShockStats}
We focus now on the statistics of shock waves as a function of the magnetic parameter $v_A$. To count the number of shock waves, an arbitrary thresholding criterion on the first-order difference signal $\delta\eta$, is fixed to $\delta\eta>5\sigma_{\delta\eta}$, with $\sigma_{\delta\eta}=\sqrt{\overline{\delta\eta^2}}$ its standard deviation. This criterion thus selects if and when a peak within the signal corresponds to a shock wave. The shock rate $\Gamma$ is defined as the average number of shocks found per second and $dt_S$ the time between two successive shocks. The number of shocks depends on $v_A$ as well as on the forcing strength quantified by the measured steepness $\epsilon$. Note that if the forcing is too weak, no shock can emerge because of viscous effects, even at high $v_A$.

Figure~\ref{DiagPhase}(a) displays the phase diagram in the ($\epsilon$, $v_A$) parameter space of the gravity-capillary wave turbulence regime and the shock wave regime. The transition between the two regimes is shown at a chosen shock rate of $\Gamma\approx0.1$~s$^{-1}$ (see the dashed line). Moreover, we observe that $\Gamma$ increases with $\epsilon$ and $v_A$, as expected, and that no shock appears, even for strong forcing, when $v_A$ is small enough ($v_A<0.25$ m/s). For these low $v_A$, stronger forcing (not achievable in our setup) would probably end up in wave breaking instead of shocks. Figure~\ref{DiagPhase}(b) shows the probability distribution function  (PDF) of the time lag $dt_S$ for all $v_A$ values and $\Gamma>0.1$~s$^{-1}$, i.e., the shock-wave regime. The PDF is independent of $v_A$ and $\Gamma$ and decreases exponentially, meaning that the shock waves are, as expected, independent and random events.

Let us now look at the probability distribution of the amplitude of the shock wave, e.g., those occurring in Figs.~\ref{wavelet_signal_0}(a) and \ref{wavelet_signal_0}(c). To do so, we compute the probability density functions of the first $\delta\eta$ and second-order difference $\delta^{(2)}\eta$ of the shock-wave amplitude for different $v_A$ as shown in Fig.~\ref{pdf_grad}. For low enough $v_A$, the distributions remain roughly Gaussian, whereas for high enough $v_A$ they follow well-defined power-law tails. It is worth noting that the power-law tail appears only for $v_A\geq0.3$~m/s, as for the occurrence of shock waves (see Fig.~\ref{DiagPhase}). The power-law tail clearly converges to $\delta\eta^{-4.3}$ for the first-order difference and to $\delta^{(2)}\eta^{-3.0}$ for the second-order difference at high $v_A$. 

A power-law tail distribution of the first-order difference is predicted in the case of diluted shocks driven by the 1D random-forced-driven Burgers equation~\cite{Chekhlov1995,weinan1997,E1999,Bec2007,Frisch2001}.  The prediction of the power-law exponent is more controversial and depends in particular on the forcing correlation degree~\cite{Frisch2001}. The PDF tail is predicted either in $\delta\eta^{-4}$~\cite{Chekhlov1995} for finite viscosity or in $\delta\eta^{-7/2}$~\cite{weinan1997,E1999} in the limit of vanishing viscosity. The observation in Fig.~\ref{pdf_grad} of a power-law PDF for the first- and second-order difference thus confirms that the shock waves drive the wave spectrum scaling. The above prediction of the power-law exponents is close to the experimental one ($-4.3$). The deviation is probably due to viscous dissipation and the fact that the experimental shocks do not generate a vertical front with a discontinuity of order one. To our knowledge, the statistics of random shock waves involving second-order discontinuities, i.e., $\delta^{(2)}\eta$ is a Dirac-$\delta$ distribution, has not been addressed theoretically but would be of primary interest to compare with our experimental results.

\section{Shock-wave spectrum}
\label{DisccusSpectra}
We have experimentally observed in Sec.~\ref{PowerSpectrum} that the power spectrum $S_\eta(\omega)$ scales as $\omega^{-4}$ when it is dominated by second-order singularities. Let us now investigate the dependence of the spectrum $S_\eta(\omega)$ with the other parameters. To derive analytically the spectrum, we follow the model of acoustic turbulence~\cite{Saffman71} but for second-order singularities, as Kuznetsov did for pointlike surface singularities~\cite{KuznetsovJETP2004}. If we assume that the second-order difference of the signal $\delta^{(2)}\eta$ is only made of a set of $N$ Dirac singularities, of amplitudes $\Delta_2(\eta)$, located at the random times $t=t_S$ ($N$ is the total number of shocks and $t_S$ are the moments they appear), one has
\begin{equation}
    \frac{\partial^{2}\eta}{\partial t^2}=\sum_{t=0}^{\mathcal{T}} \Delta_2(\eta) \delta(t-t_S)/dt^2,
    \label{Def_Dirac}
\end{equation}
where $\Delta_2[\eta(t)]\equiv \eta(t+2dt)-2\eta(t+dt)+\eta(t)$ is the second-order difference amplitude, $dt=1/f_e$, and $\delta$ is the Dirac operator. Using the Fourier transform of the surface elevation $\eta(t)$ as $\hat{\eta}_\omega=\int_{0}^{\mathcal{T}} \eta(t) e^{i2\pi ft} dt$, performing two integrations by parts to include $\partial^{2}\eta/\partial t^2$ in the Fourier transform, and then using Eq.~\eqref{Def_Dirac} and the definition of the spectrum $S_\eta(\omega)\equiv |\widehat{\eta}(\omega)|^2/\mathcal{T}$, we thus obtain
\begin{equation}
    S_\eta(\omega)=C_S\overline{\Delta_2^2}\Gamma\omega^{-4}/dt^2\ {\rm ,}
    \label{Spectrum_shock}
\end{equation}
with $C_S=1$ and the shock rate $\Gamma=N/\mathcal{T}=1/\overline{dt_S}$ with $dt_S$ the time between two successive shocks. The shock-wave spectrum of Eq.~\eqref{Spectrum_shock} thus predicts a $\omega^{-4}$ scaling (as experimentally found above), is proportional to the number of shocks, $N$, and to the variance of their amplitude  $\overline{\Delta_2^2}$, and is independent of $v_A$. Note that the acoustic spectrum of shock waves of first-order singularities scales as $\Gamma \overline{\Delta_1^2}\omega^{-2}$, with $\Delta_1\equiv\eta(t+dt)-\eta(t)$~\cite{Saffman71,KuznetsovJETP2004,KuznetsovJPP08}. More generally, for singularities of order $n$, their power spectrum reads $S_\eta(\omega)=C_S\overline{\Delta_n^2}\Gamma\omega^{-2n}/dt^{2(n-1)}$, showing that the higher $n$ is, the denser the shocks have to be to dominate in the spectrum.

\begin{figure}[t]
    \centering
    \includegraphics[width=0.96\linewidth]{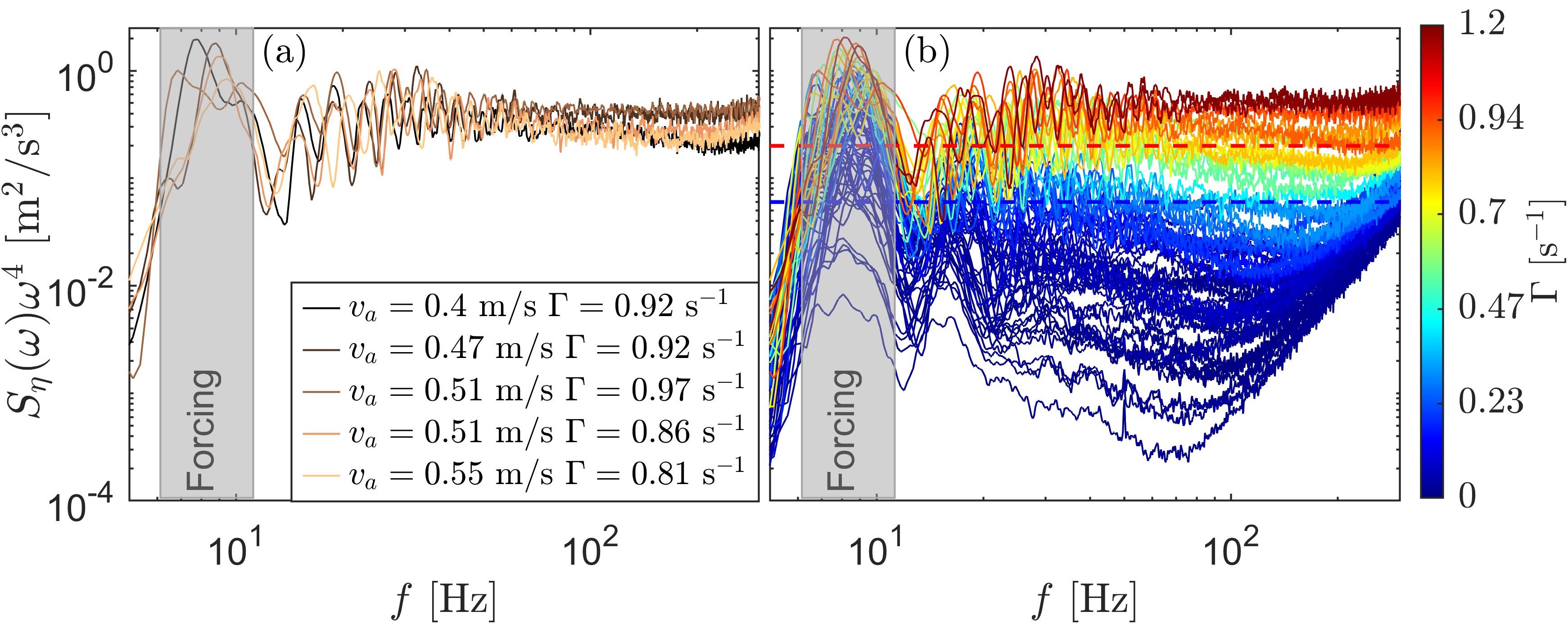}
    \caption{Frequency compensated spectra $S_\eta(\omega)\omega^4$ for (a) different $v_A$ and almost constant shock rate $0.8<\Gamma<1$ s$^{-1}$ and (b) different values of $\Gamma \in[0,1.2]$ s$^{-1}$ (i.e., $v_A\in[0,0.55]$~m/s) on a log-log plot. The horizontal blue (red) dashed lines separate the wave turbulence from the intermediate (resp., full shock wave) regimes. The gray area is the frequency bandwidth of the random forcing.}
    \label{Somegxf4}
\end{figure}

\begin{figure}[t]
    \centering
    \includegraphics[width=0.49\linewidth]{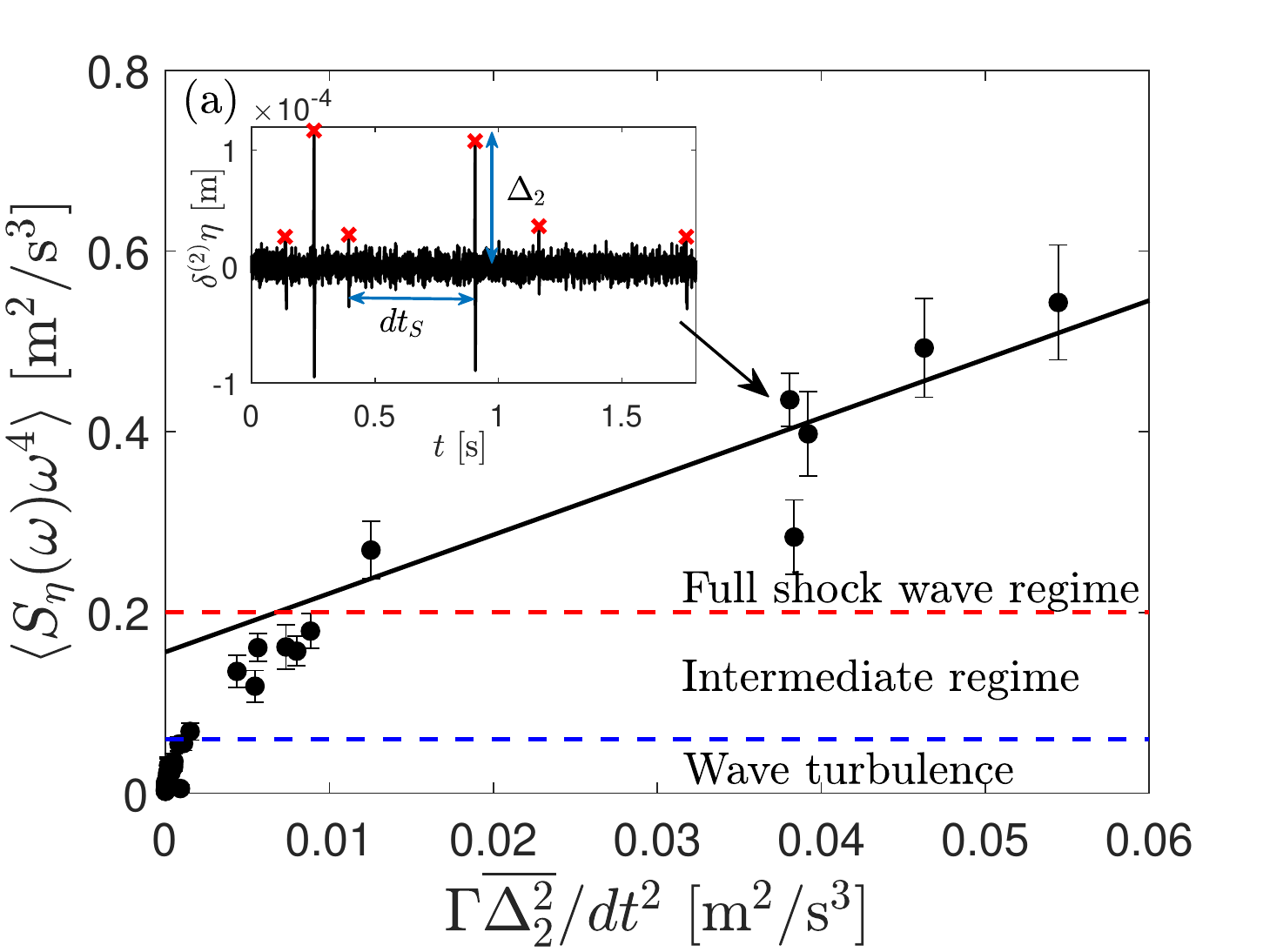} 
    \hfill
    \includegraphics[width=0.49\linewidth]{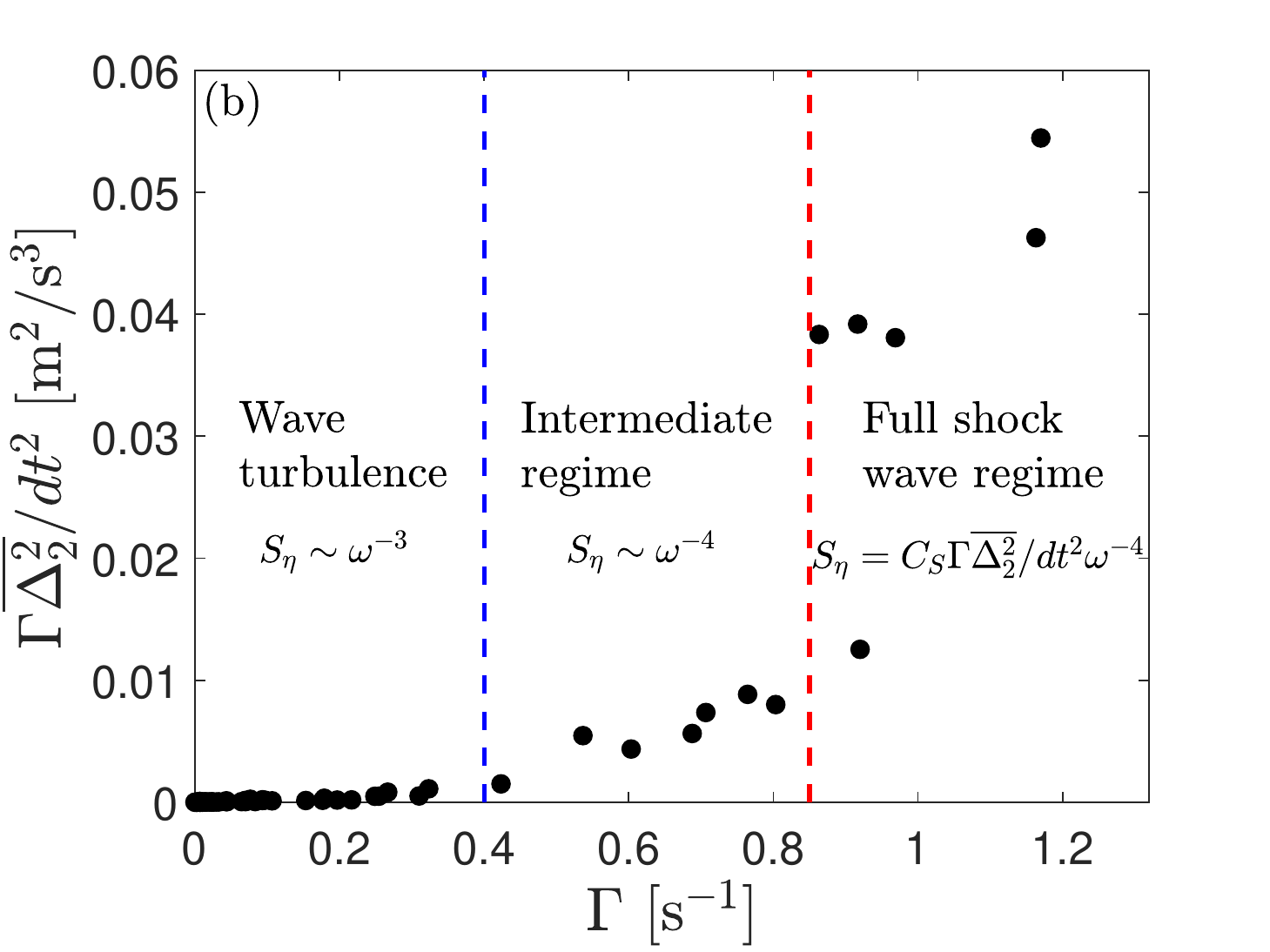} 
    \caption{(a) Compensated spectrum $\langle S_\eta(\omega) \omega^4\rangle_f$ (averaged within $80<f<170$~Hz) as a function of $\Gamma\overline{\Delta_2^2}/dt^2$. The error bars are for the standard deviation of $S_\eta(\omega) \omega^4$. The black line shows the best linear fit for $\langle S_\eta(\omega)\omega^4\rangle_f>0.2$~m$^2$/s$^3$. The inset shows the temporal evolution of the second-order difference $\delta^{(2)}\eta(t)$ (black line) and the corresponding Dirac-peak maxima (red crosses) for $v_A=0.51$~m/s and $\Gamma=0.97$~s$^{-1}$. The resulting point of the main figure is marked with an arrow. (b) Evolution of $\Gamma\overline{\Delta_2^2}/dt^2$ with the shock rate $\Gamma$. The transitions between wave turbulence, intermediate, and full shock-wave regimes are shown (see the blue and red dashed lines).}
    \label{Somegf4_GammaDelta}
\end{figure}

To test the prediction of Eq.~\eqref{Spectrum_shock}, we compute experimentally the second-order difference of $\eta(t)$ taking only the shock waves into account, i.e., we keep the maxima of the detected shock-wave events and remove the residual noise coming from the regular waves (see the red crosses in the inset of Fig.~\ref{Somegf4_GammaDelta}), the notation for $\Delta_2$ is not changed in the following, for the sake of clarity. Figure~\ref{Somegxf4}(a) then shows the experimental compensated spectrum $S_\eta(\omega)\omega^4$, which is found to be constant over almost one decade in frequency and independent of $v_A$, as expected from Eq.~\eqref{Spectrum_shock}, for a roughly constant shock rate $\Gamma$. This independence is of paramount interest and contrasts with the weak wave turbulence case in which the energy cascade is strongly dependent on the dispersion relationship. In the shock-wave regime, only the singularities and their statistics drive the spectrum once $v_A$ is high enough ($v_A>0.4$~m/s). The increase of the experimental compensated spectrum with $\Gamma$ is displayed in Fig.~\ref{Somegxf4}(b). It clearly shows that the shock rate $\Gamma$ drives the value of the spectrum amplitude. When $\Gamma>0.4$~s$^{-1}$, the scaling in $\omega^{-4}$ is achieved [see the flat compensated spectra above the horizontal blue dashed line in Fig.~\ref{Somegxf4}(b)].


Looking experimentally at the scaling of the spectrum with the shock rate $\Gamma$ and the variance of their amplitude $\overline{\Delta_2^2}$ is more challenging since fixing their values independently is not possible. However, Fig.~\ref{Somegf4_GammaDelta}(a) shows that the value of the compensated spectrum $\langle S_\eta(\omega) \omega^4\rangle_f$, averaged within $80<f<170$~Hz, increases linearly with $\Gamma\overline{\Delta_2^2}$, as expected from Eq.~\eqref{Spectrum_shock}, when the shock-wave regime is reached, i.e., for $\Gamma>0.85$~s$^{-1}$. The theoretical spectrum of Eq.~\eqref{Spectrum_shock} is thus fully verified experimentally since the experimental constant $C_S=6.5$ is found to be of the same order of magnitude as the expected unit value. For lower shock rates ($0.4<\Gamma<0.85$~s$^{-1}$), the spectrum still scales in $\omega^{-4}$ [see Fig.~\ref{Somegxf4}(b)] but its amplitude does not follow Eq.~\eqref{Spectrum_shock}, since it corresponds to an intermediate state between the shock-wave and wave turbulence regimes [see Fig.~\ref{Somegf4_GammaDelta}(a)]. Finally, Fig.~\ref{Somegf4_GammaDelta}(b) displays the evolution of $\Gamma\overline{\Delta_2^2}/dt^2$ as a function of the shock rate $\Gamma$. At low $\Gamma$, almost no shock wave is detected and a wave turbulence regime is present. For moderate $\Gamma$, the quantity increases slightly with $\Gamma$, whereas for high $\Gamma$, it increases strongly leading to a full shock-wave regime well described by Eq.~\eqref{Spectrum_shock}. 

To sum up, we used a simple model showing very good agreement with the experiments. In particular, it explains that the random shocks drive the frequency spectrum scaling in $\omega^{-4}$, whereas the number and amplitude of shocks control the spectrum amplitude independently of the value of $v_A$. We found three different regimes depending on the shock rate value: When $\Gamma<0.4$~s$^{-1}$, the shock waves are not significant enough and gravity-capillary wave turbulence occurs [below the blue dashed lines in Figs.~\ref{Somegxf4}(b), \ref{Somegf4_GammaDelta}(a), and \ref{Somegf4_GammaDelta}(b)]; when $0.4<\Gamma<0.85$~s$^{-1}$, shock waves are significant enough to develop a spectrum of second-order singularities in $\omega^{-4}$ but not enough to get the full spectrum of Eq.~\eqref{Spectrum_shock} [between the blue and red dashed lines in Figs.~\ref{Somegxf4}(b), \ref{Somegf4_GammaDelta}(a), and \ref{Somegf4_GammaDelta}(b)]; and when $\Gamma>0.85$~s$^{-1}$, the full spectrum of discontinuities from Eq.~\eqref{Spectrum_shock} is achieved [beyond the red dashed lines in Figs.~\ref{Somegxf4}(b), \ref{Somegf4_GammaDelta}(a), and \ref{Somegf4_GammaDelta}(b)]. In the latter regime, note that despite nondispersivity, the Zakharov–Sagdeev spectrum of acoustic weak--wave turbulence~\cite{Zakharov1970}, recently observed numerically~\cite{kochurin2022}, is not achieved. Shock waves indeed prevent weak turbulence. Note also that in this regime, the spectrum depends on the shock rate $\Gamma$ and so is linked to the input power. Even if a critical balance ($\tau_l>\tau_{\mathrm{nl}}^S$) is achieved (see Fig.~\ref{TimescaleFig})~\cite{NazarenkoBook}, the spectrum obtained here does not follow the Phillips spectrum that is predicted to saturate and to be independent of the input power~\cite{phillips1958}. The agreement with a spectrum of singularities, i.e., Kuznetsov-like spectrum, rather than the Phillips spectrum is here discovered experimentally for hydrodynamics surface waves and has been also observed numerically for elastic plates~\cite{miquel2013,mordant2017}. 

\section{Conclusion}\label{Conc}
We have studied the transition from quasi-1D dispersive wave turbulence to an acoustic-like nondispersive regime. To do so, we used a magnetic fluid within a canal, subjected to an external horizontal magnetic field, to tune the dispersivity of waves on the surface of the fluid. For a low magnetic field, we recovered the classical wave turbulence regime driven by nonlinear resonant interactions~\cite{Ricard2021}. For a high enough field, shock waves occur randomly involving second-order discontinuities, i.e., the second-order difference of the wave amplitude is a Dirac $\delta$. The frequency power spectrum of this shock-wave regime is found to scale as $\omega^{-4}$ and to be proportional to the shock rate and to the variance of the shock amplitudes, provided the shock rate is high enough. These experimental findings are well captured by a Kuznetsov-like spectrum of a random Dirac-$\delta$ distribution involving second-order singularities. The transition from wave turbulence to the shock-wave regime is also evidenced by measuring the energy flux. As expected, the latter is found to be constant in the wave turbulence regime and to decrease over scales in the shock-wave regime due to the damping of shock waves storing energy at all scales. When shock waves are prevalent the timescale separation hypothesis of weak turbulence theory is no longer validated experimentally and a critical balance occurs instead. The shock-wave statistics is then studied and a phase diagram between wave turbulence and the shock-wave regime is shown as a function of the control parameters. The probability density functions of the first- and second-order differences of the surface elevation are computed and found to exhibit a power-law tail with an exponent close to the predictions of the 1D random-forced Burgers equation~\cite{Chekhlov1995,weinan1997,E1999,Bec2007,Frisch2001}. 

The observation of this shock-wave regime, discovered here for surface waves, is significant for two reasons. First, the assumption of weak turbulence theory of dispersive waves has been tested experimentally with this setup and shows that the presence of shock waves prevents the possibility to reach a wave turbulence regime. Second, the energy cascades in wave turbulence due to local resonant interactions, whereas in the shock-wave regime, the energy is mainly stored in shock waves that are coherent structures rich in the frequency domain. These singularities travel over the canal length, keeping their shapes, but are damped by viscous dissipation. Theoretical and numerical works would be of paramount interest to understand in more detail the transition reported here. It would also be significant to extend the bridge between the shock-wave regime reported here, as second-order singularities, and the 1D random-forced Burgers equation. Finally, high-order statistics could be investigated experimentally in such shock-dominated acoustic regime, in particular, to test intermittency and anomalous scalings of structure functions predicted by 1D random-forced Burgers turbulence~\cite{BouchaudPRE1995,BalkovskyPRL1997,LindborgJFM2019}. 

\begin{acknowledgments}
This work was supported by the French National Research Agency (ANR DYSTURB Project No.~ANR-17-CE30-004 and ANR SOGOOD Project No.~ANR-21-CE30-0061-04) and the Simons Foundation MPS No.~651463--Wave Turbulence (USA).
\end{acknowledgments}

\clearpage
\appendix

\section{Ferrofluid characteristics}\label{Ferro}
\begin{figure}[h!]
    \centering
    \includegraphics[width=0.5\linewidth]{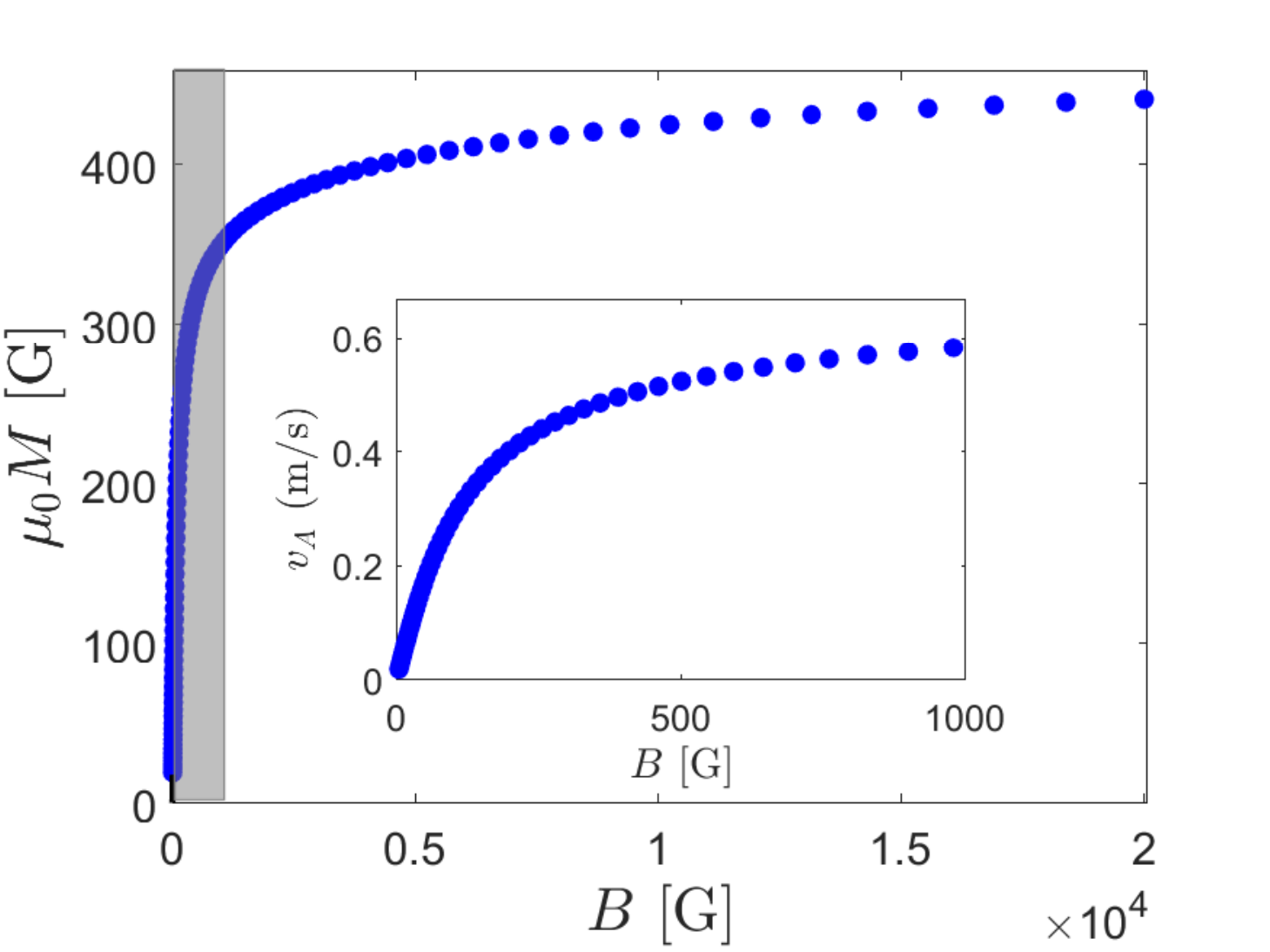} 
    \caption{Magnetization curve $M(B)$ of the PBG400 ferrofluid provided by the Ferrotec manufacturer. The gray part represents the fields achievable experimentally. The inset shows the theoretical velocity $v_A$ as a function of the applied magnetic induction $B$ corresponding to the gray part of the main figure.}
    \label{MvA_B}
\end{figure}

The magnetization curve $M(B)$ of the PBG400 ferrofluid is plotted in Fig.~\ref{MvA_B} and is provided by the Ferrotec manufacturer. It enables us to compute the variation of $v_A$ with the magnetic induction $B$ (see the inset of Fig.~\ref{MvA_B}).

\section{Time-frequency spectrum}\label{2spectra}
\begin{figure}[h!]
    \centering
    \includegraphics[width=0.7\linewidth]{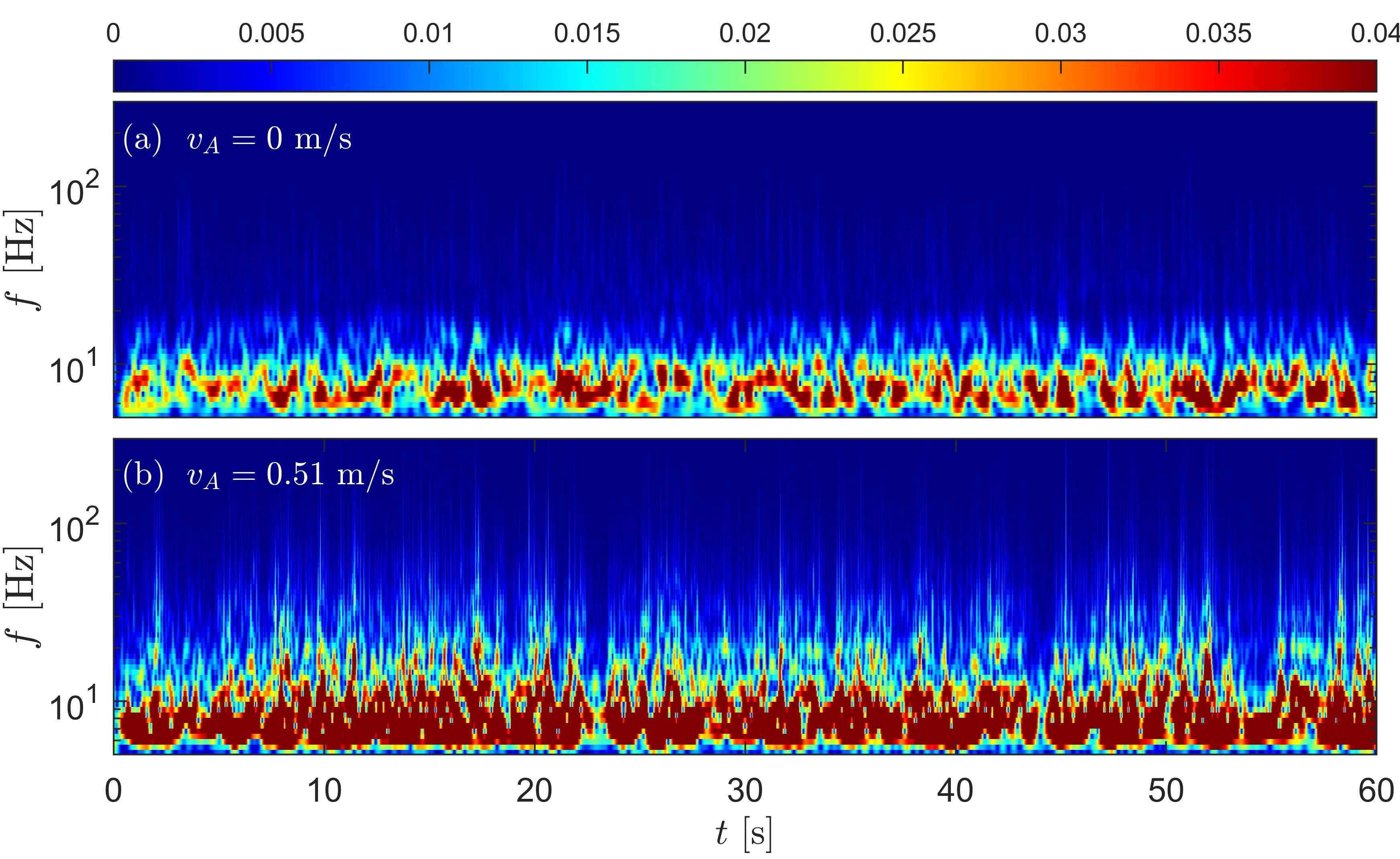} 
    \caption{Time-frequency spectrum of the surface elevation signals obtained by a wavelet transform for (a) the dispersive case ($v_A=0$ m/s) and (b) the nondispersive case ($v_A=0.51$ m/s).}
    \label{wavelet0_429}
\end{figure}

The time-frequency spectrum of the surface elevation obtained by a wavelet transform is plotted in Fig.~\ref{wavelet0_429}. In the dispersive case [Fig.~\ref{wavelet0_429}(a)], the energy cascades continuously over frequency scales and time until viscous dissipation occurs around 100~Hz. In the nondispersive case [Fig.~\ref{wavelet0_429}(b)], localized coherent structures occur randomly and contain energy to all frequency scales.

\clearpage
\section{Typical wave amplitude}\label{stdva}
\begin{figure}[h!]
    \centering
    \includegraphics[width=0.5\linewidth]{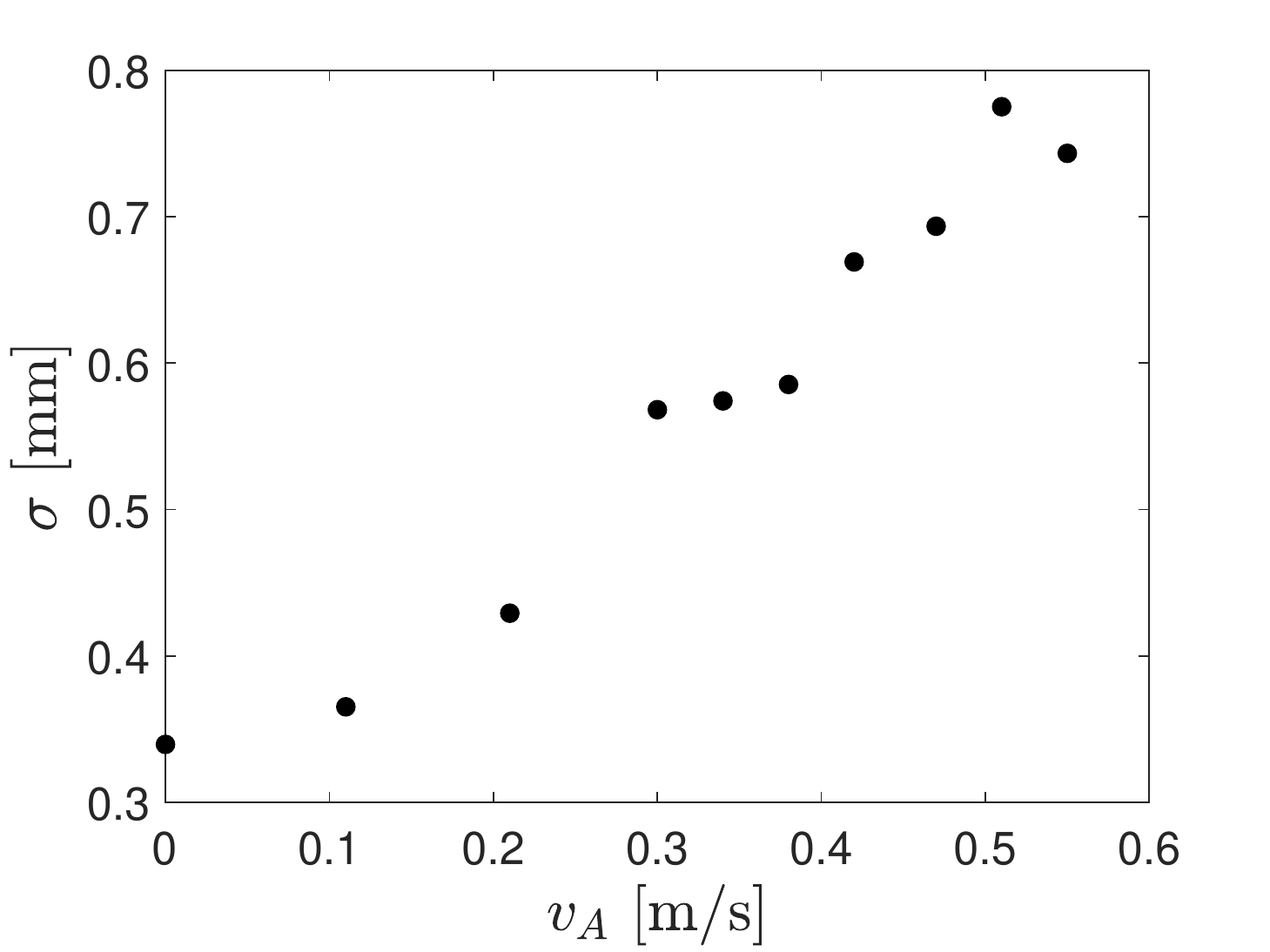}
    \caption{Evolution of the standard deviation $\sigma=\sqrt{\overline{\eta^2}}$ of the surface elevation $\eta(t)$ as a function of the magnetic parameter $v_A$ for a constant steepness $\epsilon\simeq0.07$.}
    \label{std}
\end{figure}

The evolution of the typical wave amplitude $\sigma$ with $v_A$ is plotted in Fig.~\ref{std}. It increases with $v_A$, except for the maximum value of $v_A$ where Maxwell stress due to the external magnetic field probably flattens the wave amplitude.

\section{Shock wave formation}
\label{1shock}
\begin{figure}[h!]
    \centering
    \includegraphics[width=0.575\linewidth]{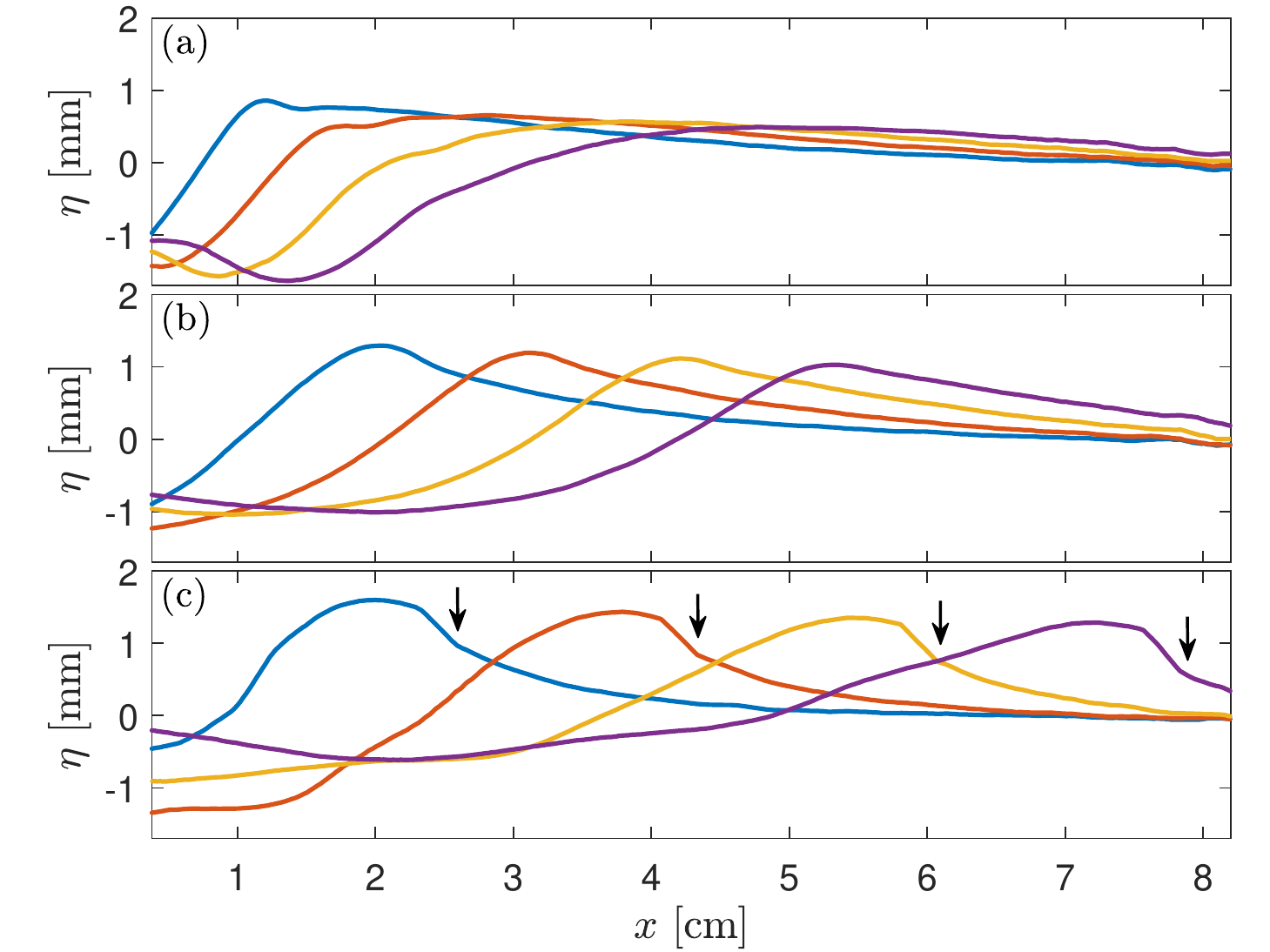} 
    \caption{Spatial evolution of a surface wave in response to a single pulse forcing for increasing times (spaced from 25 ms, from blue to purple) for the (a) dispersive ($v_A=0$~m/s), (b) intermediate ($v_A=0.3$~m/s), and (c) nondispersive ($v_A=0.51$~m/s) cases. The arrows indicate the discontinuity location over time.}
    \label{Pulse}
\end{figure}

The response of the surface to a single pulse forcing is shown in Fig.~\ref{Pulse} for different values of $v_A$. No shock wave occurs at small $v_A$ [Figs.~\ref{Pulse}(a) and \ref{Pulse}(b)] due to the dispersion. In the nondispersive case [Fig.~\ref{Pulse}(c)], a shock wave is formed and travels along the canal, keeping a constant shape with a discontinuity.

\end{document}